\documentclass[aps,prx,floatfix,reprint,showpacs,nofootinbib,superscriptaddress,longbibliography]{revtex4-2}
\usepackage{graphicx} 
\usepackage{lipsum}
\usepackage{babel}
\usepackage[normalem]{ulem}
\usepackage{xcolor}
\usepackage{bm}
\usepackage{amsmath}
\usepackage{braket}
\usepackage{amsfonts}

\usepackage[unicode=true,
 bookmarks=false,
 breaklinks=false,pdfborder={0 0 1},backref=false,colorlinks=false]
{hyperref}
\hypersetup{
 urlcolor=blue,linkcolor=black,citecolor=black
}

\begin{document} 

\title{Quantum computing for genomics: conceptual challenges and practical perspectives}

\author{Aurora Maurizio}
\email{maurizio.aurora@hsr.it}
\affiliation{Center for Omics Sciences, IRCCS San Raffaele Scientific Institute, 20132 Milan, Italy}

\author{Guglielmo Mazzola}
\email{guglielmo.mazzola@uzh.ch}
\affiliation{Department of Astrophysics, University of Zürich, Winterthurerstrasse 190, 8057 Zürich, Switzerland}

\date{\today}

\begin{abstract}
We assess the potential of quantum computing to accelerate computation of central tasks in genomics, focusing on often-neglected theoretical limitations. 
We discuss state-of-the-art challenges of quantum search, optimization, and machine learning algorithms.
Examining database search with Grover's algorithm, we show that the expected speedup vanishes under realistic assumptions. For combinatorial optimization prevalent in genomics, we discuss the limitations of theoretical complexity in practice and suggest carefully identifying problems genuinely suited for quantum acceleration.
Given the competition from excellent classical approximate solvers, quantum computing could offer a speedup in the near future only for a specific subset of hard enough tasks in assembly, gene selection, and inference. These tasks need to be characterized by core optimization problems that are particularly challenging for classical methods while requiring relatively limited variables.
We emphasize rigorous empirical validation through runtime scaling analysis to avoid misleading claims of quantum advantage.
Finally, we discuss the problem of trainability and data-loading in quantum machine learning.
This work advocates for a balanced perspective on quantum computing in genomics, guiding future research toward targeted applications and robust validation.
\end{abstract}

\maketitle


\section{Introduction}

Genomics inherently presents a suite of computationally challenging problems due to the sheer volume and complexity of biological data involved. 
The ability to accelerate these computations is of paramount importance for advancing our understanding of fundamental biological processes, disease mechanisms, and personalized medicine.\cite{alyass2015big, hasin2017multi} 
In the last decade, quantum computing,\cite{nielsen:2000} an emerging computing paradigm leveraging the principles of quantum mechanics, has garnered significant attention as a potential platform for achieving substantial speedups across various scientific and technological domains.\cite{tacchino2020quantum,cao2019quantum,di2024quantum,orus2019quantum,jayakumar2022quantum}
This naturally raises the compelling question of whether the unique capabilities of quantum computation can be effectively harnessed to address the demanding computational bottlenecks within the field of genomics. 

There is a growing body of literature discussing the prospects of quantum computation for biological and healthcare applications.\cite{outeiral2021prospects,cordier2022biology,marchetti2022quantum,10.1093/bib/bbae391,flother2023state,zinner2021quantum,emani2021quantum,ur2023quantum,maniscalco2022quantum,basu2023quantum,mcweeney2023quantum,durant2024primer,smaldone2024quantum,FLOTHER2025101236} While useful for navigating the zoo of quantum algorithms, they necessarily may lack a detailed view of genomics problems. Moreover, the theoretical and practical challenges of quantum algorithms, when applied to this realm of applications, are not always adequately addressed.

For instance, given the inherent big-data nature of many problems within genomics, it is tempting to directly assign potential quantum speedups to tasks like sequence alignment by leveraging textbook algorithms such as Grover's search.\cite{grover1996fast,jayakumar2022quantum} Similarly, considering claims regarding the computational advantage of quantum annealing hardware over classical solvers,\cite{yarkoni2022quantum,king2022coherent,king2025beyond} one might be inclined to foresee the quantum advantage across all genomics applications that rely on core combinatorial optimization routines. However, as we will discuss, these direct and perhaps overly optimistic associations often overlook crucial practical limitations and theoretical nuances that significantly impact the potential for genuine quantum acceleration in real-world genomic analyses.

The goal of this manuscript is to offer a critical perspective on the usefulness of quantum algorithms in genomics. We believe that this is particularly useful to temper some of the hype surrounding quantum computing. Practically, we hope that the concepts presented here will also be useful for researchers in planning their workflow and identifying (possible) conceptual roadblocks in their implementations.
Rather than being a compilation of works discussing the literature on \textit{quantum genomics},\cite{10.1093/bib/bbae391} we focus on the algorithms and their practical implementations.
We also discuss promising quantum sampling methods that have not been considered in previous reviews.

\begin{figure*}[ht]
    \centering
    \includegraphics[width=0.9\textwidth]{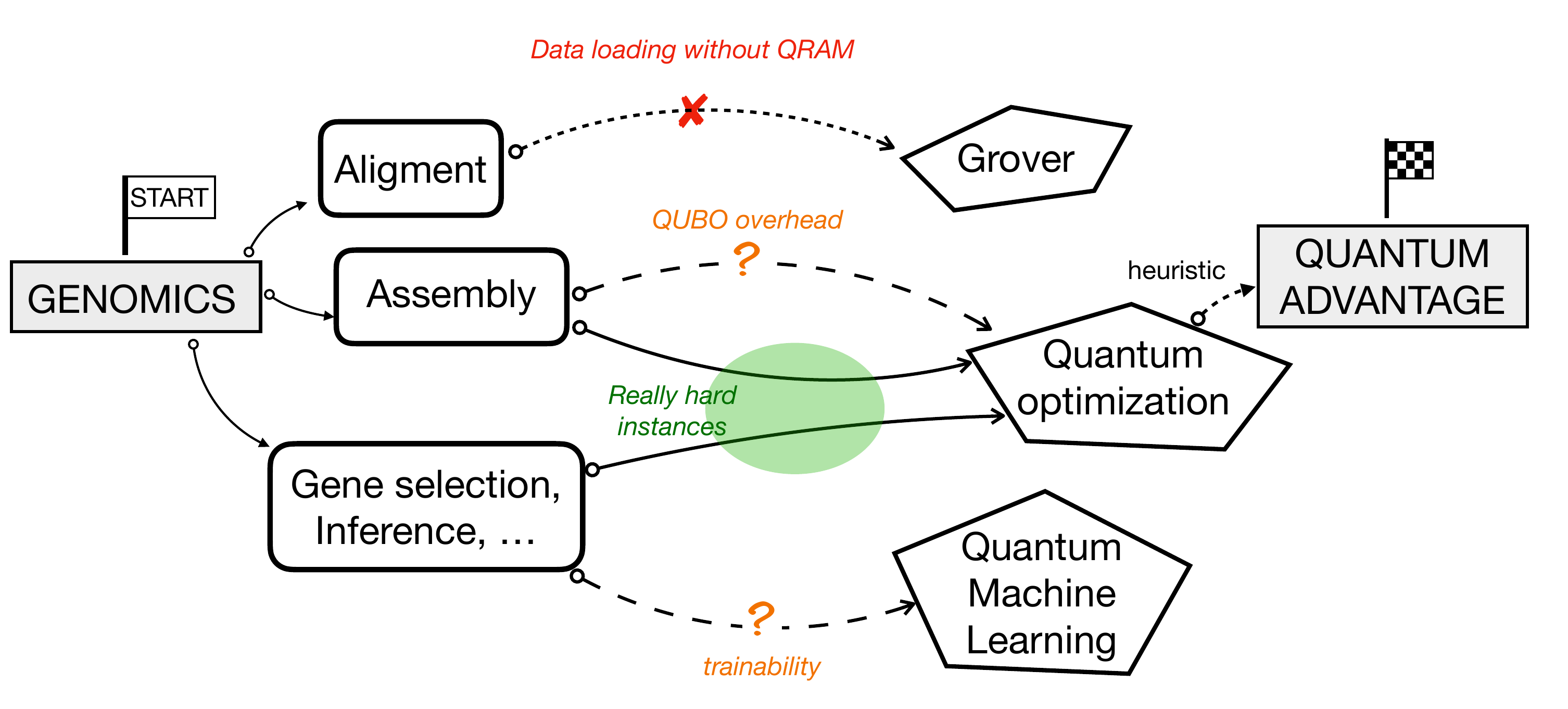}
    \caption{We schematically sketch the main connection and topics discussed in the manuscript. Concerning the genomics side, we discuss mainly three applications: alignment, assembly, and problems beyond that (here briefly labeled as gene selection and inference).
    We discuss possible quantum algorithms that have already been proposed to tackle these problems. Here, for the first time, we discuss in detail also the main conceptual and practical limitations (red, orange). We speculate that quantum advantage may be found if one identifies really hard optimization problems that do not suffer from typical overhead connected with embedding into a quantum algorithm and hardware (green pathway). 
    On the other hand, quantum search algorithms and quantum machine learning approaches may suffer from practical and conceptual limitations.
    }
    \label{fig:plan}
\end{figure*}

The conceptual map of the work is presented in Fig.~\ref{fig:plan}. The manuscript is organized as follows: In Sect.~\ref{s:introquantum}, we introduce digital quantum computing in an accessible way, along with its motivation and the definition of scaling advantage. In Sect.~\ref{s:search}, we discuss Grover's algorithm for database search, highlighting the profound challenges connected with data loading and sketching possible workarounds. In Sect.~\ref{s:optimization}, we turn to the computationally heavier task of optimization, which is inherent in many genomic tasks, from assembly to inference. We discuss the applicability of complexity theory in real-world instances and assess which optimization problems are truly hard in practice to be amenable to quantum speedup. We then discuss the challenges connected with translating a problem defined in native genomics language into a mathematical formulation suited for quantum algorithms. Moreover, we focus on heuristic methods and how to assess their scaling advantage, as well as introducing the new quantum-enhanced Monte Carlo algorithm.
Finally, we discuss current limitations of quantum machine learning in Sect.~\ref{s:qml}.
We draw our conclusions in Sect.~\ref{s:conclu}.

\section{A brief overview of quantum computing}
\label{s:introquantum}

The goal of this Section is to introduce the key concept regarding the fault-tolerant quantum computation, quantum gate frequency, and scaling advantage.
A reader knowledgeable in quantum computing may safely skip this Section. However, as a form of self-evaluation, we suggest consulting Appendix\ref{app:myths}, which presents a set of common misconceptions about quantum computation, and checking whether the arguments presented there seem trivial or not.
We present in Appendix\ref{app:basics}  the ingredients and concepts of quantum computation in the most self-contained and practical manner possible.

\subsection{Fault-tolerant quantum computation}

In this manuscript, we mostly focus on digital, fault-tolerant quantum computation,\cite{nielsen:2000} i.e., assuming no hardware noise. We will adopt the convention that a qubit is a logical qubit, not a physical one, subject to decoherence and noise.\cite{cai2023quantum} The first experimental demonstrations of error correction are currently being established,\cite{krinner2022realizing,postler2022demonstration,google2023suppressing,acharya2024quantum} so it is possible that fault-tolerant digital quantum computers will be achieved in the future.
Generally, it is expected that to fabricate one logical qubit, several physical qubits need to be bundled, with a set of operations designed to detect and correct errors.~\cite{nielsen:2000,fowler2012surface,litinski2019game}
Two promising methods to achieve error correction in superconducting qubit platforms are the ``surface'' code (pursued by Google)\cite{,google2023suppressing,acharya2024quantum} and the ``bicycle'' code (pursued by IBM)\cite{bravyi2024high,yoder2025tour}.
In short, the surface code requires local physical qubit connectivity, and a sustain a high error threshold at the cost of requiring many physical qubit to encode a logical one. 
This means that algorithms requiring only hundreds of logical qubits will, in turn, necessitate millions of physical qubits.\cite{beverland2022assessing} 
Bicycle codes, on the other hand, aim for much lower qubit overhead and potentially better performance, often at the expense of more complex qubit connectivity and decoding.

In our view, current demonstrations using noisy machines should be regarded solely as algorithmic demonstrations and proof of principles, as they are unlikely to scale effectively for executing algorithms with provable quantum advantage.
However, the possibility of achieving quantum advantage with heuristic algorithms and noisy analog machines,\cite{preskill2018quantum} perhaps in specific models and instances, is not ruled out. This could be the case for optimization\cite{abbas2024challenges} or sampling.\cite{mazzola2024quantum}
However, here we will mostly consider fault-tolerant quantum computation because: (i) it represents the most realistic pathway to quantum advantage; (ii) it is conceptually easier to understand quantum algorithms within this framework; and (iii) it is hardware agnostic, allowing us to discuss algorithms at a high level without the need for modeling the microscopic quantum processes (e.g. quantum tunneling, decoherence, etc.) occurring in the hardware.\cite{das2008colloquium,albash2018adiabatic}

\subsection{Scaling advantage vs gate time}

There are two kinds of quantum algorithms\cite{montanaro2016quantum,jayakumar2022quantum}: those with a provable scaling advantage over their best-performing classical counterparts, and those that offer only a heuristic speed-up, i.e., that cannot be demonstrated mathematically. A scaling advantage means that the runtime of the quantum algorithm, 
$T_Q$, scales with the system size $N$ of the problem in a more favorable manner than the runtime of a classical algorithm, $T_C$.

\begin{figure}[t]
    \centering
    \includegraphics[width=0.9\columnwidth]{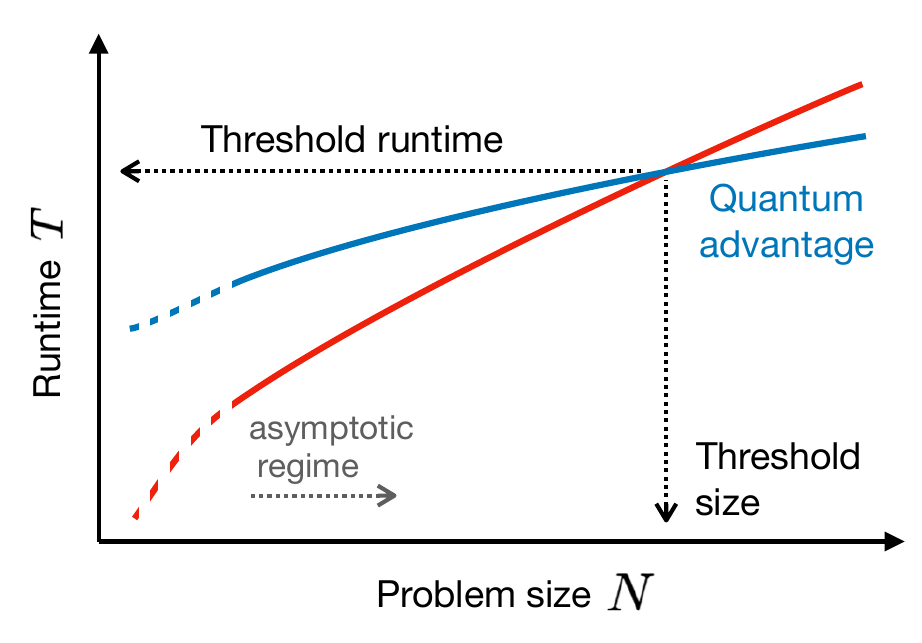}
    \caption{Illustrative example of the scaling of runtimes $T$ for quantum (blue line) and classical (red line) algorithms in a case where quantum speed-up is possible. In the asymptotic regime, i.e., for sufficiently large sizes, the two runtimes scale differently with $N$. However, due to the significantly longer gate times for quantum computing, the two lines cross only at a threshold problem size. This implies a threshold runtime. Some applications may involve typical problem sizes that are much smaller than this threshold. In such cases, a practical quantum advantage might only emerge at an unrealistically large problem size.
    }
    \label{fig:runtime}
\end{figure}

Let's consider, as an example, the task of database search over a list of $N$ elements, the worst-case performance for the conventional approach requires 
$N$ lookups, while Grover's algorithm only requires $\sqrt{N}$
  lookups, realizing an appealing quadratic speedup. However, there are two potential issues to consider in more detail: (a) the quantum database lookup must be efficient and should not scale with $N$ in a way that negates the speed-up, and (b) even if the lookup can be performed with a number of operations independent of $N$ (in jargon, a constant-depth oracle), thereby retaining the full quadratic speed-up, the fault-tolerant quantum logic gates operate at a much slower clock frequency compared to conventional computers.\cite{gidney2019efficient,babbush:2021, beverland2022assessing,mazzola2024quantum}

The typical quantum gate frequency will depend on the architecture and error correction code implemented in the future, but estimates suggest around 10 kHz\cite{gidney2019efficient}. Given that the execution of a quantum gate requires several control cycles, a quantum gate cannot be faster than the classical electronics used to control it. It is accepted that the quantum runtime will be affected by a much larger prefactor due to this significant gate frequency mismatch. This implies that quantum advantage, if any, will only occur for sufficiently large system sizes. The size at which the quantum algorithm outperforms the classical one is referred to as the threshold size. This situation is shown in Fig.~\ref{fig:runtime}, and discussed in the context of genome search in Sect.~\ref{ss:runtimegrover}.

To conclude, the existence of a provable quantum advantage is not always sufficient to predict practical quantum advantage in real-life use cases. A polynomial speed-up can be overshadowed by the large prefactor due to gate times, revealing itself only for problem sizes that are unrealistically large and therefore not practical. Even exponential speed-ups, where the threshold size is expected to be smaller, can be negated by other factors, such as the data-loading problem,\cite{Aaronson:2015scy,tang2021quantum} which will be discussed in the next section.

\section{Quantum search in genomics}
\label{s:search}

\subsection{Adapting the Grover algorithm for DNA alignment}

The idea of obtaining a quadratic speed-up for search problems, using the Grover algorithm, is particularly appealing in applications such as sequencing and alignment. The first mention of such applications dates back to 2000,\cite{bioinfo2000Hollenberg} with the initial use of the phrase \emph{quantum bioinformatics}. The task involves finding the location of a target sequence (or string) within a database. Examples include searching within the human DNA, where the string represents a sequence of DNA bases (A, T, G, C), or at higher levels, a sequence of amino acids, and so on.
Several works propose a quantum algorithm solution for the alignment problem\cite{prousalis2019alpha,niroula2021quantum,sarkar2021qibam,soni2021quantum,miyamoto2023quantum}.
However, in this Section, we present arguments questioning the real scalability of such methods in a practical setting.

  \subsubsection{Data encoding}

First, we need to consider how to encode our problem-specific variables into binary form. For instance, in the task of finding a string made of $M$ DNA bases, such as ATGAAC..., within a larger DNA segment of $N$
elements, we need a way to represent the four DNA bases using binary encoding. Since each DNA base can take one of four values, we require two qubits to encode this information. We can assign the following convention: A = '00', T = '01', G = '10', C = '11'. Similarly, other data types can be encoded in a similar way. For example, encoding amino acids would require 5 bits since there are 20 different amino acids, and $2^4 < 20 < 2^5$.

\begin{figure*}[ht]
    \centering
    \includegraphics[width=0.9\textwidth]{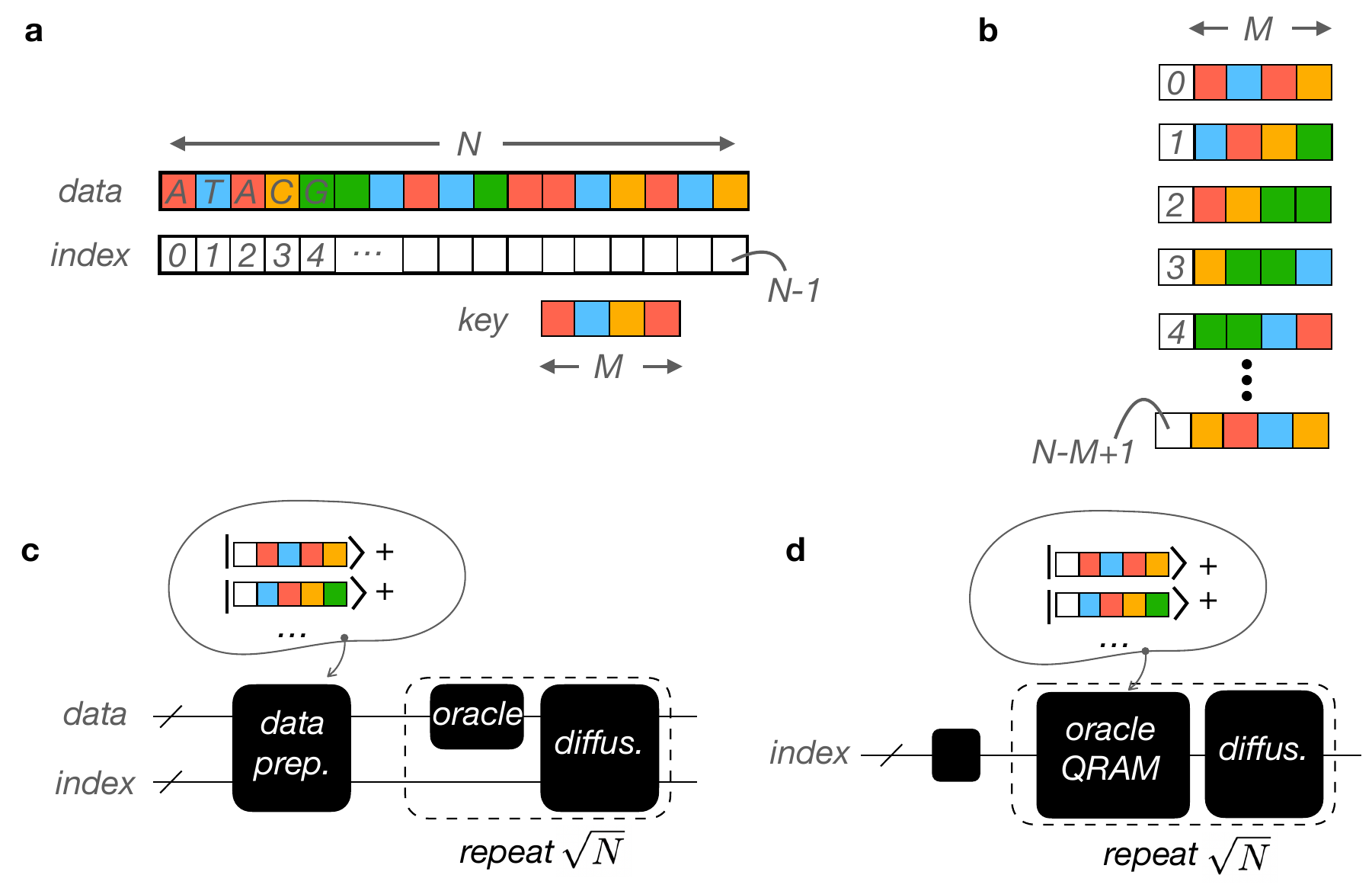}
    \caption{DNA reads alignment using the Grover algorithm. (a) Sketch of the DNA portion of length $N$. Each position is labelled by an index. The task is to find the `key' string of length $M$. (b) The problem is translated into finding the `key' element in a database of $N-M+1$ strings of length $M$. (c) Grover algorithm implementation that does not require QRAM. It features a \emph{index} register, made of $log_2(N)$ qubits, and a \emph{data} register, made of $2M$ qubits. The entangled state that encodes the full ordered database is created using a data preparation unitary, which requires $N$ operations to encode the data exactly. (d) Grover algorithm implementation with QRAM. In this case, the task of loading the data in superposition is offloaded to the Grover oracle. 
    }
    \label{fig:genome}
\end{figure*}

In DNA reads alignment, it is not only necessary to find whether a pattern exists in the database but also to determine its location in the genome. The full array of data to search from is thus the labeled set of 
$N-M+1$ of all possible strings of length 
$M$, inside the full genome of length 
$N$ (see Fig.~\ref{fig:genome}).

At this point, there are basically two possible directions in which the Grover algorithm can be adapted to this problem. The first one already admits a circuit implementation, while the second one follows the textbook route of simply invoking a black-box oracle and offloading all the complexity of querying the database to this operation.

\subsubsection{Grover without QRAM}

We will start with the first route, which we believe is particularly instructive in unraveling the hidden issues of data loading. This method was originally proposed in Ref~\cite{bioinfo2000Hollenberg}, and also adapted in Ref.~\cite{sarkar2021qibam}.

The core idea of this approach is to entangle the qubit register of the indexes with the qubit register of the data.
The index can be encoded in binary form, therefore requiring $q_i = \log_2{(N)}$ qubits (assuming $N$ being a power of $2$ for simplicity).
The data register requires $q_d = 2 M$ qubits, assuming the encoding above relating DNA elements and bits.
Since $N\gg M$ this encoding can be considered pretty efficient memory-wise. 
For the human genome case, $N \approx 3 \times 10^9$ such that $\log_2{(N)} \approx 32$, while typical reads lenght are in the range $M\sim50-100$.
The total number of logical qubits needed to encode the ordered database is $\log_2{(N)} +2M$, so approximately about $200$ logical qubits.

This way to encode the data and the indexes together provides a practical way to construct the circuit for the Grover oracle, at the expense of using marginally more qubits, but more importantly, a costly operation for creating the entangled initial state. A toy model of this problem is provided in the Appendix~\ref{app:groverexample}. 
Unfortunately, given that the genome data does not follow any deterministic pattern, the depth of the circuit needed to load the data cannot scale more favorably than $N$.

We conclude this subsection by noting that the data-loading problem is one of the major hurdles to overcome in quantum algorithms for user-generated data. A somewhat similar situation arises in linear algebra applications. In this case, the HHL quantum algorithm\cite{harrow2009quantum} solves a system of $n$ equations in 
$log(n)$ steps, thus realizing an exponential speed-up compared to the classical method, which scales as $n^3$.
However, the exponential advantage vanishes if one considers that on the order of $n^2$
  elements need to be loaded in some way. 
The usual way forward is to assume the existence of a quantum RAM (QRAM),\cite{giovannetti2008quantum,Arunachalam_2015} which is a procedure capable of loading a bunch of classical data into superposition. 
In some rare cases, it can be argued that the $n^2$
elements can be auto-generated using a much smaller number of inputs, which can then be efficiently loaded.\cite{clader2013preconditioned} However, this seems hardly the case, to say the least, for the human genome dataset.

\subsubsection{Grover with QRAM}
 The second approach to genome search using Grover's algorithm is to utilize QRAM\cite{giovannetti2008quantum} within the oracle.
Given two registers, $|i\rangle_{\textrm{index}} | 0\rangle_{\textrm{data}}$ as before, one first create a superposition of all addresses (this is inexpensive, requiring just one layer of Hadamard gates for each of $log_2(N)$ qubits),
$ 1/\sqrt{N} \sum_{i=0}^{N-1} |i\rangle_{\textrm{index}} | 0\rangle_{\textrm{data}}$, and then obtain the superposition of values
\begin{equation}
     \frac{1}{\sqrt{N}} \sum_{i=0}^{N-1} |i\rangle_{\textrm{index}} | 0\rangle_{\textrm{data}}  \xrightarrow[]{\text{QRAM}}  \frac{1}{\sqrt{N}}\sum_{i=0}^{N-1}  |i\rangle_{\textrm{index}} |d_i\rangle_{\textrm{data}} .
\end{equation}
The marked entry is then identified within the oracle, and a minus sign is applied to the corresponding index. The rest of the Grover algorithm proceeds as described in textbooks.

The crucial weak point of this black box procedure is that it is still under debate whether QRAM can be experimentally realized, even in the distant future.\cite{Arunachalam_2015}
For a detailed explanation of the conceptual and experimental challenges of QRAM, we refer to the recent review\cite{jaques2023qram}.

\subsection{Possible solutions and approximate matching of DNA reads}

This issue of the cost of loading the database in the Grover algorithm is known\cite{emani2021quantum} but has been mostly overlooked in
literature aiming to apply Grover's algorithm (or its variants) to biological data. 
In Ref.~\cite{tezuka2022grover}, this problem is examined in the context of pattern matching in images. A possible solution proposed there involves using a parameterized low-depth quantum circuit. This could enable approximate amplitude encoding\cite{nakaji2022approximate} and might be realized using a purely variational circuit or a generative adversarial network.
The parameters in the variational circuit could be obtained just once through an expensive learning or optimization process, but could then be transferable. 
Given that the core primary sequence of the human reference genome is fundamentally stable and major updates occur only every few years, \cite{NCBI_Human_GRC,es2001initial,venter2001sequence} this idea could, in principle, make sense. However, there are no theoretical bounds on the resulting errors. 
It is not even clear that effective variational training can be performed if the target distribution lacks structure, as the optimization of parameters could encounter barren plateaus\cite{larocca2024review} in the parameter space, preventing meaningful optimization. This issue could be further exacerbated by quantum measurement noise.\cite{scriva2024challenges}
Another solution has been proposed in Ref.~\cite{niroula2021quantum} which however, incurs having a qubit register of size $N$, namely exponentially larger than before. 

One last ray of hope, concerning the theoretical applicability of Grover's algorithm in DNA reads matching, may arise if we consider that biological data contains errors. These errors could reflect true biological events (e.g. genomic alterations such as polymorphisms and somatic mutations) or technical artifacts introduced during sample preparation, and sequencing. 
Common types of artifacts affecting sequenced reads include substitutions, where one nucleotide is incorrectly read as another; insertions, where extra nucleotides are erroneously added; and deletions, where nucleotides are skipped or lost.
Most of the time, the key string to be searched contains these errors, meaning the real practical task is not exactly as defined above, but rather to find the closest match instead of the exact match. High stringency (i.e., low error tolerance) may lead to low alignment rates by discarding too much data, while low stringency (high tolerance) can compromise accuracy by allowing incorrect matches. 

This occurrence has been considered in the literature,\cite{sarkar2021qibam} however, the proposed solutions still suffer from the inefficient data loading explained above.

Given the unavoidable errors in the data, an approximate amplitude encoding protocol~\cite{tezuka2022grover} could, in principle, be viable. However, it is not possible to make a conclusive statement as the fidelity of the prepared state using shallow circuits is difficult to predict. Small-size demonstrations, i.e., fewer than 20 qubits, are unlikely to be predictive of the quality of the state preparation required for the 200 qubit case.

\subsection{Constraining quantum runtimes for a simple alignment task}
\label{ss:runtimegrover}

Besides the theoretical challenges behind QRAM, to achieve a practical quantum speed-up in DNA reads alignment applications, one needs to carefully assess the target runtimes. 
To be more specific, let us assume, for the sake of argument, that it is indeed possible to overcome the previous barriers related to the data loading problem. We are interested in understanding if a quadratic speed-up would still provide a meaningful advantage, given the realistic hardware limitations of future quantum computers.

Before proceeding, some clarifications are in order. This manuscript aims to be accessible to both the quantum computing and bioinformatics communities, acknowledging that there may be differing definitions of `runtime' for the tasks discussed. Most bioinformatics tools typically address the full problem of alignment, inherently dealing with factors such as sequencing errors.
On a typical high-performance computing setup with multiple CPU cores (e.g., 16-32 cores) and sufficient memory (e.g., 64-128 GB RAM), the alignment process for a whole genome sequencing of a human sample at 120X coverage and 100bp paired end reads using e.g. Burrows-Wheeler Aligner might take anywhere from a few hours to a day, depending on the available resources and system load.

In contrast, if one considers the idealized core problem of simply finding an exact string of length $M$ within a database of length $N$, the classical runtime is significantly shorter. This specific sub-problem is the only component that a quantum search algorithm is designed to accelerate. Therefore, a meaningful comparison of quantum and classical performance should be confined to this sub-task.
Again, modern bioinformatics tools do not perform a naive linear scan of the entire genome. They employ highly optimized string-matching algorithms and data structures. 
For the task of exact matching a $M=$ 100bp string over the whole human genome, $N = 3 \times 10^9$ bp long, these tools may take less than 1 minute.
A quadratic speed-up implies a number of `oracle' calls in Grover's algorithm, which is $N_{\textrm{calls}} = \sqrt{N}  \approx 6\times 10^4$.
We therefore ask the total `quantum' runtime, $T_Q = t_\textrm{call} \times N_{\textrm{calls}} $, where $t_\textrm{call}$ is the physical runtime to execute one oracle call (see e.g Fig~\ref{fig:examples_grover}), to be shorter than 1 minute.
This yields a $t_\textrm{call}=$1 ms time to execute one step of Grover's algorithm.
This constraint arises simply from the assumed quadratic speed-up and the known runtime of classical methods. 

One can then make assumptions about the logical gate time of quantum hardware to determine whether this maximum time-per-step can be met.
For a logical gate frequency of 10 kHz, quantum advantage would only be achievable if a single step of the Grover oracle consisted of an unrealistically short circuit with a depth of 10 (given that 1/(10 kHz) $\times$ 10  = 1 ms). Conversely, an error-correcting architecture that could enable a much more optimistic logical gate frequency of 10 MHz would permit circuits as deep as 10.000 layers.

These sorts of order-of-magnitude estimates are crucial for quickly gauging the real-world feasibility of a quantum algorithm under practical conditions. They offer a rapid sense of whether a theoretical speed-up could genuinely translate into an advantage on actual hardware.

\section{Optimization problems in genomics and quantum advantage}
\label{s:optimization}

\subsection{Optimization problems in genomics}

Hopes of achieving quantum advantage are more realistic for harder computational problems in genomics, which are connected with combinatorial optimization.
Many computational problems in genomics can be recast as optimization problems. Below, we provide some popular examples.

\subsubsection{Sequence alignment}

DNA sequencing is becoming faster and cheaper at a rate far beyond Moore's Law. Recent technological advancements and a dramatic decrease in sequencing costs have led to an unprecedented increase in sequencing throughput. Modern sequencing technology can now generate millions of reads from a sample, leading to an explosion of biological data that presents significant computational challenges.\cite{chikhi2016compacting} 
Traditional computer systems often struggle to handle this data explosion due to fundamental limitations in storage, processing power, memory capacity, and data transfer rates. An example is the Von Neumann bottleneck, where computing performance is constrained by the continuous data movement between the memory and the CPU. However, a crucial question we aim to address, and partly answer, in this manuscript is whether some computational slowdowns can be resolved by quantum computing.
To explore this, we need to better define what it means for a problem to be \textit{challenging}. For many practitioners, a problem may become difficult when it can no longer fit into available memory. For instance, aligning a large dataset against a comprehensive reference database is a routine, computationally challenging bioinformatics task that might require tens to hundreds of gigabytes of RAM. This memory demand can quickly exceed the capabilities of standard desktop machines, necessitating access to high-performance computing clusters or cloud-based solutions, as well as memory-efficient algorithms and optimized software pipelines.

However, as we showed earlier, memory access can also be a bottleneck for quantum algorithms. Therefore, we expect that a practical quantum advantage is more likely for combinatorial problems defined on a relatively small input set $N$, but for which the time to solution grows exponentially, i.e., $\exp(N)$, or at least with a very steep polynomial degree.

Not all optimization problems fit this description. For instance \textit{pairwise sequence alignment} using dynamic programming algorithms, such as Needleman-Wunsch\cite{needleman1970general} and Smith-Waterman,\cite{smith1981identification} has a time and space complexity of $\mathcal{O}(NM)$, where 
$N$
  and 
$M$ are the lengths of the two sequences. 
Moreover, there exist heuristic methods such as the Basic Local Alignment Search Tool (BLAST) that achieve almost linear scaling.\cite{altschul1990basic}
Further, these algorithm can be accelerated by using of special-purpose architectures such as GPUs or FPGAs.
Therefore, we do not expect an advantage from quantum computing in this case.\cite{babbush:2021,hoefler2023disentangling}

\textit{Multiple sequence alignment (MSA)} is instead much more difficult and known to be an NP-hard problem, meaning that the computational time required to find the \textit{exact} optimal solution grows exponentially with the number and length of the sequences.
This would make it a good candidate for quantum advantage.
However, approximate methods exist, and even if they are not guaranteed to yield the optimal alignment, they are widely used in practice for tasks such as de novo assembly.
These methods have a hidden combinatorial complexity, which is overcome by the usage of heuristic solutions inside the pipeline.
For instance, the overlap-layout-consensus (OLC)\cite{idury1995new,schatz2010assembly,miller2010assembly,flicek2009sense} involves comparing all pairs of reads to find overlaps.
However, the layout phase, which involves finding a Hamiltonian path in the overlap graph, is NP-hard. Therefore, OLC methods typically rely on greedy approaches and heuristics to find a good, but not necessarily optimal, assembly.

An alternative approach is given by graph-based algorithms, where the assembly problem is transformed into finding specific paths within these graphs.
Specifically, the \textit{de Bruijn graphs} method is particularly successful for assembling the large numbers of short reads.\cite{li2008mapping,iqbal2012novo} It works by first breaking the sequencing reads into shorter subsequences of a fixed length $k$, called $k$-mers. Each unique $k$-mer is then represented as a node in a directed graph. An edge is drawn from one $k$-mer to another if the suffix of the first $k$-mer of length $k-1$ is the same as the prefix of the second $k$-mer of length $k-1$, effectively representing an overlap of $k-1$ bases. The genome assembly problem is then transformed into finding a path that visits every edge exactly once (called Eulerian path).
The Eulerian path problem is generally linear in the number of edges, so this approach is preferred for handling large numbers of short reads. 
Essentially, OLC and De Bruijn graph-based approaches are preferred depending on the context. OLC is typically used for longer reads, while De Bruijn graphs are more suitable for shorter reads.
Therefore, quantum computing should focus on problems where these approximate, yet highly effective, classical methods are insufficient to fully complete the task.
For instance, $k$-mers graph are less efficient in dealing with repetitive regions and sequencing errors.

This problem is also connected with finding graph similarities in a network. For this task, quantum algorithms based on Hamiltonian simulation\cite{daskin2014multiple}, or support vector machines equipped with graph kernels have been proposed\cite{kishi2022graph}.

\subsubsection{Genome-wide association studies}

Computationally hard problems in genomics extend beyond assembly tasks. For instance, \textit{genome-wide association studies} (GWAS)\cite{tam2019benefits,uffelmann2021genome} involve complex search and optimization tasks,\cite{moore2010bioinformatics} as they aim to identify genetic single-nucleotide polymorphisms (SNPs) that are associated with diseases.

\subsubsection{Phylogenetic inference}

Another genomics subfield of interest is \textit{phylogenetic inference}, also known as phylogenetic tree reconstruction. This is the process of inferring the evolutionary relationships between a set of biological entities (e.g., species, genes) based on their DNA, RNA, or protein sequences.\cite{schwartz2017evolution,rannala2008phylogenetic} 
The core optimization problem in phylogenetic inference lies in searching the vast space of possible tree topologies (the branching patterns) and branch lengths (representing the amount of evolutionary change) to find the tree that best explains the observed DNA, RNA data under a specific model of sequence evolution. The number of possible unrooted binary trees grows exponentially with the number of leaves, making an exhaustive search computationally infeasible. Again, powerful heuristic search strategies are applied, these include approximate maximum likelihood,\cite{chor2005maximum} branch and bound,\cite{hendy1982branch} distance-based methods.\cite{trees1987neighbor}

\subsubsection{Haplotype phasing}

Finally, NP-hard optimization problems arise in the \textit{haplotype phasing} problem.\cite{browning2011haplotype}
This is the task of determining the specific combination of alleles (i.e., variants of a gene) on the chromosome copies.
The optimization problem in haplotype phasing is determining the most likely phase configuration for each heterozygous variant along the chromosome. This is often done using sequencing reads that span multiple heterozygous sites, allowing researchers to determine which alleles are inherited together on the same DNA strand. However, the number of possible haplotype combinations grows exponentially with the number of heterozygous sites.
Interestingly, this problem can be recast as a \textit{Max-Cut} problem, which is well known in optimization theory.\cite{ding2001min,festa2002randomized}
This approach formulates the original problem as a graph partitioning problem. A fragment graph is constructed where the vertices represent the alleles observed at heterozygous sites in the sequencing reads. An edge is placed between two vertices (alleles at different sites) if they are observed together on at least one sequencing read. The weight of the edge reflects whether the co-occurrence is consistent with the same haplotype (positive weight) or different haplotypes (negative weight). The problem then becomes finding a cut in the graph that separates the vertices into two sets (representing the two haplotypes) such that the total weight of the edges crossing the cut is maximized (or, equivalently, the weight of edges within each partition is maximized).
This NP-hard problem is then solved approximately\cite{yu2021spechap}, including Markov Chain Monte Carlo methods\cite{browning2007rapid,browning2007rapid}
Note that the Max-Cut problem is one of the most commonly used models to test new optimization schemes. Due to its straightforward mapping to the Ising Hamiltonian,\cite{lucas2014ising} it is also widely employed in quantum optimization studies.\cite{crooks2018performance,guerreschi2019qaoa,wang2018quantum,wurtz2021maxcut}

\subsection{Complexity theory classes vs practical problems}

\subsubsection{NP-hard problems}

For practical purposes, even an algorithm with quadratic scaling, such as the one for pairwise alignment, can present significant challenges when 
$N$ is very large, even though such a problem is considered ``easy'' from the standpoint of complexity theory due to its polynomial scaling. As discussed, quantum methods are expected to demonstrate competitive advantages earlier in problems that are harder than polynomial.\cite{babbush:2021}
Here we discuss what kind of advantage is expected.
From a purely theoretical point of view, an exponential quantum advantage is not expected for optimization problems belonging to the NPO-hard class (strictly speaking, the NP-hard label refers to the decision problem; the corresponding optimization problem should be labeled as NPO-hard. However, in the following, we will adopt the common abuse of language and use the NP-hard label for both.) However, this does not exclude the possibility of large speed-ups in practical problems.
Indeed, complexity theory typically deals with \emph{worst-case} scenarios. In practical applications, one deals mainly with \emph{average-case} instances. This means that, despite theoretical results that indicate that there is no quantum advantage for NP-hard problems in the worst case, quantum computers could still achieve exponential speedups in some cases.\cite{abbas2024challenges,jordan2024optimization}

Moreover, we suggest that many problems in genomics also defy conventional complexity classes since (i) we may be interested in reasonably good solutions rather than the exact one (see below), and (ii) the input data itself can contain experimental errors, such that the exact solution for the slightly misdefined problem is not expected to be exact for the 'true' data.
The presence of errors may favor one method over another. For instance, sequencing errors degrade the performance of the De Bruijn graph method. Therefore, in realistic scenarios, a candidate quantum algorithm must demonstrate robustness against expected classes of errors.

This strongly supports the use of heuristic algorithms, which, while lacking formal guarantees, often strike an excellent balance between cost and accuracy. The potential benefits of such algorithms must be assessed empirically, depending not only on the problem class but also on the specific encoding used. Moreover, it is crucial to consider the current and foreseeable limitations of quantum hardware.\cite{abbas2024challenges}

Finally, we notice that in genomic problems, simply finding the \textit{fastest} solution isn't enough, and algorithms that sacrifice biological interpretability for raw speed might be less valuable in a biological context. 

\subsubsection{Polynomial time approximations}

Use cases in genomics should be carefully chosen in advance to enhance the likelihood of achieving a potential quantum advantage.
 Indeed, not all NPO-hard optimization problems are challenging in the same way, in practice. Most known complexity theory results hold for the task of finding the exact solution of a problem, but in practice, if we relax this goal to simply find a good approximation, then the same problem may belong to a different, `easier' class.\cite{abbas2024challenges}
For instance, an optimization problem belongs to the PTAS (polynomial-time approximation scheme) class if there is a polynomial-time algorithm that finds a solution whose value is within a distance $\epsilon$ from the true optimum.\cite{ausiello2012complexity} Some types of problems, like the Traveling Salesman Problem (TSP), whose general formulation is NP-hard, may indeed change their complexity class in PTAS if more structure (e.g Euclidean distances) is injected into their formulation.\cite{arkin1994approximation} Practice also shows that the TSP can be approximately solved for up to millions of nodes.\cite{abbas2024challenges}

Suitable problems are those that (i) present significant challenges for all existing methods and (ii) become difficult even with a relatively small number of constituents. This ensures that the asymptotic regime is reached within the current capabilities of the hardware, allowing for the extraction of meaningful empirical quantum runtime scaling that can be compared to classical approaches.

\subsection{From genomics to optimization and back}

In this Section, we aim to briefly review previous attempts at solving problems in genomics using quantum optimization (`bottom-up'), and to propose a new `top-down' pathway for connecting the two disciplines.

\subsubsection{Bottom-up studies}

So far, quantum computing has been proposed for solving optimization problems arising from genomic tasks, such as sequence alignment, \cite{matsumoto2022qualign} de novo assembly,\cite{sarkar2021quaser,boev2021genome,nalkecz2022algorithm,chen2023vrp} phylogenetic trees,\cite{dinneen2023qubo,park2024scalable} transcription factor-DNA binding,\cite{li2018quantum} epistasis analysis,\cite{hoffmann2024network} codon optimization,\cite{fox2021mrna} densest k-subgraph,\cite{calude2020quantum} and secondary structure prediction of mRNA sequences.\cite{kumar2025towards}
The common strategy is to map these problems into Quadratic Unconstrained Binary Optimization (QUBO) problems, which, being defined in terms of binary variables, are well-suited for implementation in quantum software and hardware.\cite{lucas2014ising}

QUBOs correspond to Hamiltonians that can be implemented on quantum annealers and have been the subject of empirical studies for more than a decade.\cite{ronnow2014defining,mandra2016strengths,denchev2016computational,boixo2014evidence} However, QUBO problems can also be executed on digital quantum devices and tailored to algorithms such as the Quantum Approximate Optimization Algorithm (QAOA).\cite{farhi2020quantumapproximateoptimizationalgorithm}
The QUBO method requires minimizing the following cost function
\begin{equation}
    H(x) = \sum_{i} h_i x_i + \sum_{i < j} J_{ij} x_i x_j,
\end{equation}
where $x_i$ are binary variables with values in $\pm 1$, and $h_i, J_{ij}$ are system dependent couplings.
A real-world optimization problem, defined using its `native' variables, weights, and constraints, needs to be recast in this format.\cite{glover2018tutorial}

Most of these studies take a ``bottom-up'' approach, directly seeking evidence of quantum speedup for use cases of interest. Unfortunately, in order to fit the sizes and connectivity constraints of current quantum hardware, the original problems are often simplified or divided into smaller subproblems. This raises concerns about whether these processed models retain the same complexity as the larger, original problem.
While the merit of these studies lies in demonstrating how real-world problems can be -in principle- mapped onto a given quantum platform, they usually do not prove or disprove a quantum advantage. In reality, even empirical demonstrations of scaling advantages require careful numerical experiments, making the task highly nontrivial (see Sect.~\ref{ss:qaoa}).

\begin{figure*}[ht]
    \centering
    \includegraphics[width=0.9\textwidth]{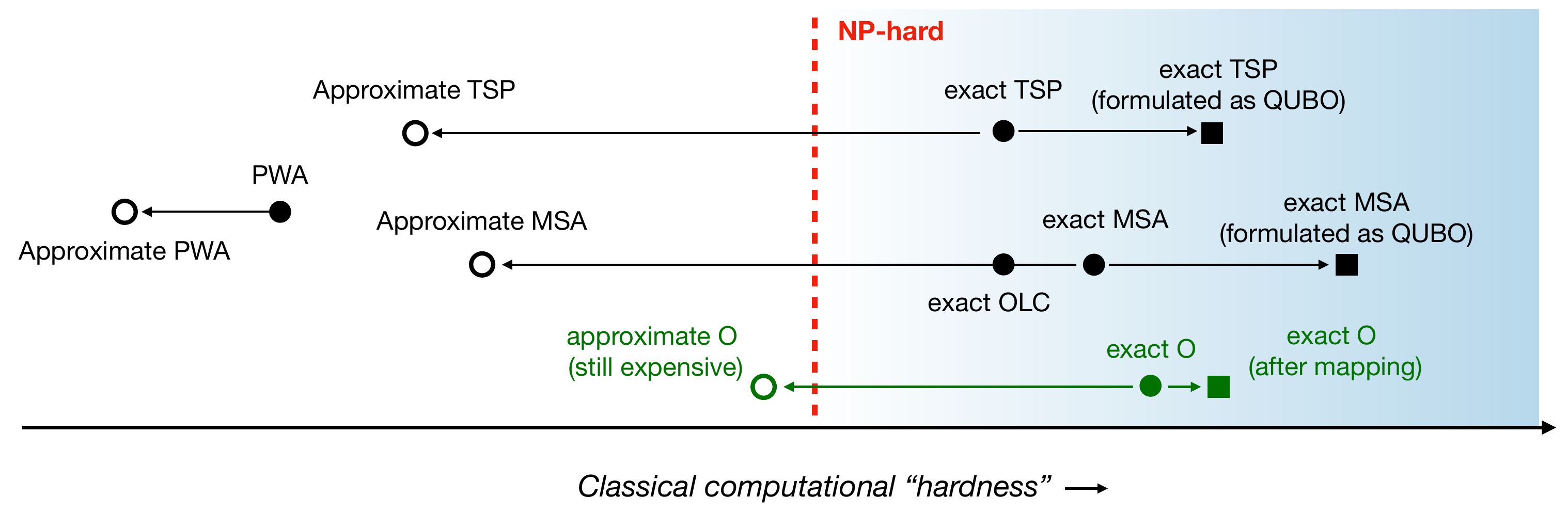}
    \caption{Optimization problems in genomics, classical hardness, and approximations. 
    We sketch the comparative computational hardness of several optimization problems discussed in the text, including pairwise alignment (PWA), multiple sequence alignment (MSA), and the traveling salesman problem (TSP). The exact formulations (indicated with solid circles) of MSA and TSP  belong to the NP-hard class (whose boundary is drawn as dashed-red line), while PWA can be solved in polynomial time. All these problems can also be approached with approximate methods (indicated with empty circles), which trade accuracy for speed. For example, PWA can be solved with nearly linear scaling using BLAST, and the overlap-layout-consensus (OLC) problem, while still theoretically NP-hard, can also be solved approximately in practice. We further illustrate that these native NP-hard problems can be translated into QUBO form (solid squares), typically incurring a polynomial overhead. This implies that a quantum algorithm must solve a version of the problem that is harder than the original formulation. Lastly, we sketch the likely necessary conditions for an generic optimization problem (labeled as ``O'')  to be amenable to quantum advantage (green symbols): (i) all known classical approximations remain computationally challenging or perform poorly in terms of accuracy, and (ii) the problem is already formulated in a way that minimizes the overhead associated with quantum embedding.
    }
    \label{fig:hardness}
\end{figure*}

\subsubsection{Practical limitations of QUBO formalism}

Here we discuss an essential \textit{limitation of QUBO method}, which is often overlooked, or rarely explicitly discussed in applied works.
While in theory every NP-hard problem can be reformulated as another with only polynomial overhead,\cite{lucas2014ising,nannicini2019performance} this polynomial factor can become a practical bottleneck. In practice, mapping an original optimization problem into a QUBO often introduces significant memory and time overhead.\cite{glover2018tutorial} This is particularly true when the original problem includes constraints that rule out large regions of the computational space, as these would encode unphysical or invalid solutions, a situation that almost always occurs in genomics problems such as assembly.

For example, the TSP, defined over $N$ cities (nodes), requires $N^2$ variables (hence qubits) when mapped into a QUBO formulation.\cite{lucas2014ising,nannicini2019performance} The same quadratic overhead affects existing QUBO-based approaches to genome assembly.\cite{sarkar2021quaser,boev2021genome} Therefore, solving the TSP using its `native' formulation is much more efficient than doing so via a QUBO mapping.

While complexity theory considers polynomial overhead negligible for NP-hard problems, in practice, where powerful classical heuristics exist, such overheads may make a critical difference.
In Fig.~\ref{fig:hardness} we qualitatively illustrate this scenario.

Another significant challenge arises when applying the QUBO problem to quantum machines with limited connectivity, such as current quantum annealers. Due to the sparse connectivity of the hardware, a mapping of the original problem onto the machine is necessary for fully connected QUBO problems. In other words, the qubits in the system are not linked to every other qubit but are only coupled with a subset. Embedding fully connected graphs introduces another quadratic overhead, which compounds the previously mentioned overhead. This can be particularly frustrating since the all-to-all connectivity requirement stems from the QUBO formulation itself, rather than the `native' problem! For instance, any optimization problem involving constraints on a sparse graph will turn into an all-to-all problem when translated into a QUBO form. To handle this artificial all-to-all structure on real quantum hardware, the problem must be embedded, resulting in an additional quadratic overhead.\cite{konz2021embedding}
This could lead to a quantum slow-down rather than a speed-up, until new quantum annealer architectures with long-range qubit connections are developed.

However, there are also counter-examples where the QUBO formulation provides an enhancement already at the problem-formulation level.
For a lattice protein design task, the resulting QUBO model solved on classical machines outperforms previous classical state of the art-solvers.\cite{panizza2024protein}

\subsubsection{Top-down studies}

An alternative, ``top-down'', strategy could be to first identify which general classes of optimization problems are particularly amenable to quantum speedup and then determine whether such problems appear in genomic pipelines.

As of today, there is no widely accepted benchmark set of problems for evaluating progress in quantum optimization. In Abbas et al.~\cite{abbas2024challenges}, a portion of the community working in this area attempted to define such benchmarks. The proposed criteria are:
(i) The problems must be empirically known to be sufficiently hard even at limited sizes.
(ii) The problems must be embeddable in binary models, and (iii) approximate classical solvers perform comparatively worse in these cases.
Examples directly quoted in Ref.~\cite{abbas2024challenges} include the Quadratic Assignment Problem, the Knapsack problem, the Maximum Independent Set problem, Low Autocorrelation Binary Sequences, Sports Timetabling Problems, and spin glasses.
The fact that these problems become hard at relatively small sizes enables a reliable scaling analysis of runtimes. Specifically, there are no computational phase transitions as the system size increases that could disrupt the evaluation of scaling exponents.
Recently, a library featuring instances belonging to these classes have been published.\cite{koch2025quantum}

\subsubsection{Possible genomics applications of very hard optimization problems}

In the following, we review some of the known applications of these classes of problems in the realm of genomics, at least to the best of our knowledge. We also speculate on relations with genomics problems when specific literature is missing.

\textit{Quadratic Assignment Problem (QAP).}
The mathematical formulation of a QAP typically involves minimizing the sum over all pairs of nodes $i$ and $j$ of the product of the `flow' between them, $w_{ij}$, and a generalized `distance' between them $d(f(i),f(j))$, where $f(i)$ is a function acting on each node.\cite{burkard1997qaplib} 
The QAP does not have an approximation algorithm running in polynomial time, making it a good candidate for quantum advantage. Since QAP models problems where the cost of an assignment is determined by the relationships between pairs of entities, they seem relevant for genomic problems.
However, the use of QAP in genomics seems so far to be limited to enhancing visualization of microarray data based on gene expression similarity,\cite{inostroza2007integrated,inostroza2011qapgrid}, potentially revealing intricate relationships in gene expression data that might not be readily apparent through more localized clustering methods.

\textit{Knapsack Problem.}
The Knapsack problem is a well-known optimization problem that involves selecting a subset of items, each characterized by a weight and a value, such that the total weight of the chosen items does not exceed a given capacity, while simultaneously maximizing the total value of the selected items.
To the best of our search capabilities, we did not find any direct application of the Knapsack problem in the genomics literature. However, we attempt to draw some connections to gene selection.
Gene selection is a critical task involving the identification of a subset of genes that are most pertinent to a particular biological inquiry.\cite{guyon2002gene,diaz2006gene}
This includes identifying genes associated with a specific disease, genes that play a key role in a biological pathway, or genes that can effectively classify different biological samples (e.g. during the experimental design for a custom gene sequencing panel). Practical constraints and the goal of selecting the most `valuable' genes make the Knapsack problem a potentially suitable framework for modeling this challenge.

For instance, in gene panel design, the value of a gene might correspond to its clinical relevance, mutation frequency, or predictive power in distinguishing disease subtypes. 

The weight associated with selecting a gene could reflect experimental costs, technical difficulty, or a penalty for redundancy, such as including genes that are highly correlated with others in the panel in the set, aiming to promote diversity.
The capacity in this context may be defined by constraints such as fixed panel size, budget limit, or sequencing coverage, all of which impose a limit on how many and which genes can be included.
Framing gene panel optimization as a Knapsack problem thus allows for a formal approach to balancing utility and constraints in a principled manner.

\textit{Maximum Independent Set (MIS).}
An independent set in a graph is defined as a collection of vertices wherein no two vertices are adjacent. The MIS problem seeks to identify such a set with the largest possible number of vertices.
The structure of the MIS problem, which centers on the selection of non-interacting elements, shares similarities with several challenges within genomics. Many problems in genomics involve the identification of sets of entities, such as genetic markers that are mutually incompatible or non-overlapping in some sense.
Interestingly, MIS problems can be mapped to sparse QUBO ones.\cite{abbas2024challenges}
Some problems, such as identifying unrelated individuals for genetic analysis, have been recast as a MIS\cite{staples2013utilizing,abraham2014identifying}
This is of particular importance in
GWAS, where a fundamental assumption for many statistical tests is the independence of the samples being analyzed.  However, human populations often exhibit complex patterns of relatedness, which, if not accounted for, can lead to spurious associations between genetic variants and the traits or diseases under investigation. The MIS problem provides an elegant solution to identify the largest possible subset of individuals within a dataset who are statistically considered `unrelated'. In this application, the first step involves calculating the pairwise relationship between all individuals in the study. This is typically done using genetic data, such as SNPs, to estimate measures like identity by state or identity by descent. Based on these data a graph is constructed where each individual is represented as a vertex. An edge is drawn between two individuals if their estimated relatedness exceeds a predefined threshold, indicating that they are considered `related' (based on a threshold) for the purposes of the study. Then the problem of identifying the largest set of unrelated individuals becomes equivalent to finding the maximum independent set in this graph.

Further, MIS has also been proposed in gene selection. By modeling the redundancy between genes as edges in a graph, one can use the MIS framework to select a subset of genes that are maximally informative and non-overlapping in their biological roles.\cite{bull2013maximising}
Finally, the problem of identifying non-repetitive promoters has been recast as an MIS problem in Ref.~\cite{hossain2020automated}.

\subsection{Heuristic methods for quantum optimization}
\label{ss:qaoa}
As of today, quantum optimization demonstrations feature two main strategies: quantum annealing (QA)\cite{johnson2011quantum,das2008colloquium}, implemented on analog quantum hardware, and the Quantum Approximate Optimization Algorithm (QAOA)\cite{farhi2014quantum,Zhou2020} or similar approaches based on parameterized circuits, like the Variational Quantum Eigensolver (VQE)\cite{peruzzo2014}, executed on digital hardware. We refer the reader to reviews of these algorithms for definitions and details\cite{mcclean2016,cerezo2021variational,das2008colloquium,albash2018adiabatic}.

\begin{figure*}[ht]
    \centering
    \includegraphics[width=0.9\textwidth]{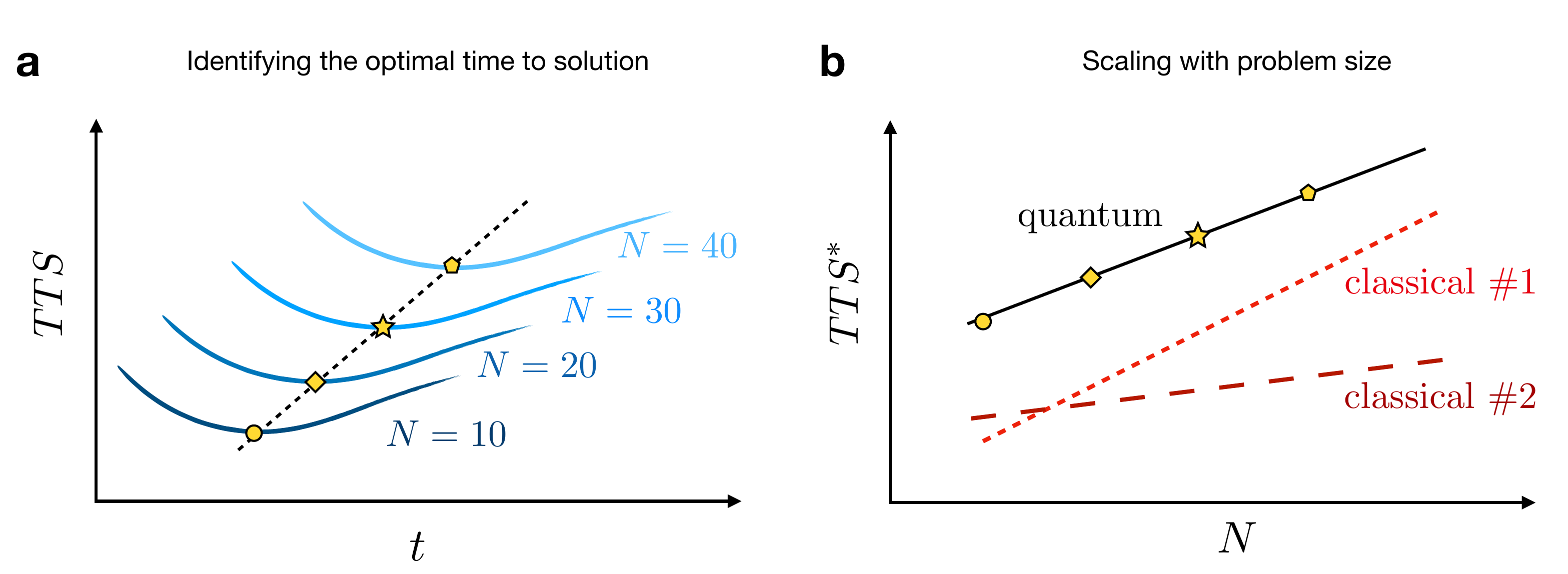}
    \caption{Sketch of an empirical assessment of scaling advantage: \textit{(a)} Identifying the optimal time to solution,  $TTS^*$ as a function of the time $t$ of the run. For each problem size $N$ (in this hypothetical scenario $N=10, 20, 30, 40$) this represents the minimal cost of the simulation. \textit{(b)} Scaling of the  $TTS^*$ as a function of $N$. This plot is used to assess the existence of an empirical scaling advantage over classical methods.
    In this sketch, the quantum algorithms would outperform the first classical method, but not the second. 
    }
    \label{fig:anneal}
\end{figure*}

However, it must be noted that all these approaches are heuristic in practice. Quantum annealing yields the exact solution only in the infinite runtime limit. Moreover, analog machines are strongly affected by decoherence, allowing coherent dynamics to persist only for tens of  nanoseconds, see e.g.\cite{king2022coherent}. Current quantum annealing experiments extend beyond tens of microsecond runtimes, leading to strong coupling between the system and the environment\cite{das2008colloquium,ronnow2014defining,albash2018demonstration}. As a result, the computational resources of these machines involve a combination of incoherent tunneling events and conventional thermal annealing\cite{boixo2014evidence,denchev2016computational}.

Quantum optimization on today’s digital hardware is similarly affected by gate noise, as these machines are not yet fault-tolerant\cite{Zhou2020,Sack2024qaoa,miessen2024benchmarking,pelofske2024scaling}. QAOA can be interpreted as a discrete version of quantum annealing, where the total evolution is divided into layers or blocks, each associated with tunable parameters\cite{farhi2014quantum}. These parameters are iteratively adjusted to minimize the cost function measured at the end of the execution.

Recent large-scale simulations of QAOA reveal that coherent dynamics are maintained up to about 8–10 layers (for a simple disordered Ising model). Beyond that point, gate noise introduces errors that increase the cost function\cite{miessen2024benchmarking}. Additionally, optimizing the variational parameters is itself a non-trivial task, even in an ideal noiseless scenario\cite{bittel2021training}.
While several strategies have been proposed to mitigate this challenge\cite{Sack2019qaoa,Egger2021warmstart,scriva2024challenges}, it is clear that QAOA also remains a heuristic method.

This means that the execution of both QA and QAOA, must be repeated multiple times in practice. In the literature, this is commonly discussed in terms of the \emph{success probability}, $p$, that is, the probability of finding the exact solution or a solution whose energy cost is below a certain threshold.

Since each run occupies the quantum hardware for a physical time $t$, the total \emph{time to solution} is $TTS= R \times t$, where $R$ is the number of repetitions required to achieve the desired success probability.
Ideally, one would perform a single execution with $t \rightarrow \infty$. However, this is obviously impractical. The opposite limit consists of performing a huge number of repetitions, each with a very short execution time $t$, but this is also inefficient, since one would perform no annealing.
It has been recognized that there exists a trade-off between the total annealing (or circuit) time $t$ and the number of repetitions $R$ required to achieve a target success probability $p$ (for example, $p = 0.90$).

Determining the optimal balance between $t$ and $R$ is essential for accurately assessing the scaling of the total computational cost with the problem size\cite{albash2018demonstration}. Notably, the \emph{ optimal time to solution}, defined as
$TTS^* = \textrm{min} (t~R(t))$ 
depends on the size of the problem, hence $TTS^* = TTS^*(N) $.
This is the proper quantity to plot against $N$ to assess the scaling of the quantum time cost vs problem size. 
This procedure is shown graphically in Fig.~\ref{fig:anneal}.
As clearly explained in the Refs.~\cite{ronnow2014defining,albash2018demonstration}, it is important to identify the optimal time to solution when analyzing the scaling. Overestimating the runtime for smaller problem sizes would result in a flatter scaling curve, potentially leading to an artificial indication of quantum advantage.

While this concept is particularly transparent in the case of QA, the existence of an optimal time (or resource usage) to solution also applies to QAOA and VQE. In these cases, the number of repetitions includes not only the number of independent executions of the runs, but also the number of times each circuit is executed, referred to as shots, to collect sufficient statistics for evaluating the cost function at each step. The total runtime 
$t$ is proportional to the circuit depth. The trade-off between the number of shots, repetitions, and circuit depth is discussed in Ref.~\cite{scriva2024challenges}.

Historically, two-dimensional spin glasses (well-suited to the architecture of quantum annealers) have been studied extensively\cite{boixo2014evidence,ronnow2014defining,Katzgraber_2015}. We refer interested readers to Ref.~\cite{ronnow2014defining} which represents a key work, providing a cautious and thorough assessment of the potential quantum advantage of annealers over classical heuristics such as simulated annealing (SA) \cite{kirkpatrick1983optimization}.
Subsequent research has attempted to engineer artificial models that could be amenable to quantum speed-up. While these models may lack direct practical relevance, they help to explore which types of energy landscapes might favor quantum annealing. Ref.~\cite{albash2018demonstration}, for example, identifies a class of artificial problems where quantum annealers demonstrate scaling advantages over SA. Interestingly, Ref.~\cite{micheletti2021polymer} extends this observation to more realistic problems related to polymer folding on a lattice, showing a similar scaling advantage.
Q

While these results are promising, they do not constitute conclusive evidence of quantum advantage. A definitive demonstration would require outperforming all possible classical solvers. For instance, Ref.~\cite{albash2018demonstration} shows that quantum annealing outperforms SA but not simulated quantum annealing\cite{santoro2002theory}, a quantum-inspired heuristic designed to approximate quantum annealing dynamics on a classical machine.

To conclude, seeking quantum advantage through empirical observation is certainly a legitimate and practical approach. However, careful attention must be paid to how the scaling with problem size is analyzed and to ensuring fair comparisons with the best-known classical methods for the same problem. Moreover, classical algorithms should be appropriately optimized for the task at hand.
For example, if the optimization problem possesses a known structure, this prior knowledge should also be exploited by the classical algorithm. A notable case concerns Ref.~\cite{denchev2016computational}, where an initially claimed scaling advantage of quantum annealing over SA was refuted by employing a slightly more sophisticated cluster-update SA algorithm\cite{mandra2016strengths}.

\subsection{Quantum enhanced Monte Carlo}
\label{ss:montecarlo}

Many genomics analyses (e.g., phylogenetic tree inference \cite{mau1999bayesian} and population genetics models \cite{maccluer1967monte}) rely heavily on Monte Carlo methods. Speeding these up with quantum algorithms could be transformative.

Besides QAOA and quantum annealing, another practical quantum approach that may be useful for optimization-related tasks, particularly sampling, has recently emerged. A novel quantum algorithm capable of accelerating Markov chain Monte Carlo was proposed and has since been demonstrated on noisy quantum hardware.
The concept was initially introduced by one of us in 2021~\cite{mazzola2021sampling}. Later, Ref.~\cite{layden2023quantum} demonstrated a polynomial speed-up in the simulation of an all-to-all connected spin glass model. Notably, this speed-up persisted despite hardware noise, although the demonstration was limited to small system sizes.

Obviously, the potential to accelerate Monte Carlo simulations could have a significant impact on a wide range of genomics tasks involving statistical inference, such as those tackled today with Hidden Markov Chain. Moreover, this approach is directly relevant to the optimization of NP-hard problems, as it could enhance the performance of simulated annealing.\cite{arai2025quantum}

However, this method is still relatively new, and its strengths and limitations\cite{orfi2024bounding,christmann2025quantum} have yet to be thoroughly evaluated before concrete applications in genomics can be expected.

Finally, quantum-enhanced sampling can also be achieved using annealers. 
For instance, they have been successfully applied to efficient QUBO models for DNA packaging.\cite{slongo2023quantum}



\section{Quantum Machine learning}
\label{s:qml}

Many tasks in life science require some form of machine learning, which might lead one to consider quantum machine learning (QML)~\cite{biamonte2017quantum, cerezo2022challenges, schuld2015introduction, du2025quantum}  as a potential candidate to speed-up several workflows.\cite{PhysRevA.111.042416,FLOTHER2025101236}
QML is a rapidly growing field, and it is outside the scope of this work to provide an in-depth discussion of its status and latest achievements. However, to provide a coherent picture of the conceptual challenges of quantum algorithms in genomics, it is worth mentioning general issues of QML.

First, it should be noted that, to date, no quantum speed-up for machine learning on \textit{real-world} datasets has been observed or even postulated. Any scalability advantages have only been demonstrated on artificial datasets.\cite{havlivcek2019supervised,liu2021rigorous} In particular, a robust quantum speed-up for QML has been suggested within the context of quantum kernel methods applied to datasets constructed from the Discrete Logarithm Problem\cite{liu2021rigorous}. It has been rigorously shown that a quantum kernel method can achieve an exponential speed-up in classification accuracy over any efficient classical learner. This advantage originates from the fact that the discrete logarithm problem is widely believed to be classically hard, while Shor's algorithm provides an exponential speed-up for solving it on a quantum computer.\cite{shor1999polynomial} This allows the quantum feature map to separate data in a way that appears random to classical algorithms.

Nevertheless, important limitations exist. First, this speed-up has only been proven for a specific, mathematically constructed dataset. Whether such an `inherently quantum-hard' structure exists in real-world, naturally occurring datasets (i.e those for which we actually want to apply machine learning) is still an open question. Second, the method relies on a fault-tolerant quantum computer capable of efficiently implementing a quantum feature map whose circuit is as demanding as performing the Shor's algorithm. Although this is theoretically valuable, it should not be seen as an immediate enabler of practical QML with current hardware.

Similarly, as in quantum optimization, we anticipate the tension between \textit{top-down} and \textit{bottom-up} strategies. One approach to demonstrate the advantage in QML for genomics would be to detect similar structures within genomic datasets and wait until quantum hardware capable of running deep circuits (as deep as the Shor algorithm) becomes available. Alternatively, an empirical approach might involve developing heuristic strategies that establish a practical advantage considering several datasets.

However, in QML, there's an additional challenge. Quantum neural networks (QNNs), or similar methods based on optimizable architectures, are significantly harder to train than their classical counterparts. In current literature, this issue is often referred to as the \textit{barren plateaus} problem, where the optimization landscape becomes exponentially difficult to optimize as system size increases.\cite{mcclean2018barren,larocca2025barren} Much of the current research aims to understand the occurrence of barren plateaus and identify ways to mitigate them.\cite{holmes2022connecting,ortiz2021entanglement} Barren plateaus do not occur if circuits are sufficiently shallow\cite{zhang2024absence} or highly structured,\cite{pesah2021absence} but this could also make these circuits easier to simulate classically, potentially negating any quantum advantage.\cite{cerezo2023does,bermejo2024quantum,Schreiber_2023}
Overall, a quantum learning model cannot demonstrate a true quantum advantage if it lacks the necessary trainability, expressivity, or capacity for generalization that surpasses its classical equivalents. 

The key caveat we want to emphasize is that, while discovering a quantum architecture that outperforms the best classical approaches is clearly important, 
one needs to take into account 
trainability issues that only emerge at larger scales.

To summarize, a QML method based on a trainable architecture will need to meet the following requirements to achieve practical quantum advantage: (i) trainability at large scale;
(ii) inability to be efficiently simulated classically (otherwise, it could be considered a quantum-inspired kernel running on classical machines);
(iii) superior performance over the best classical models on real-world datasets;
(iv) efficient input of data.

The latter issue applies broadly to quantum algorithms aiming to process generic datasets. In this context, there are generally two practical approaches to loading data into a quantum circuit for training:
The first is data encoding via angles of parameterized rotation gates.\cite{havlivcek2019supervised} Here, the properly normalized data can be embedded as rotation angles within the feature map. While this approach is widespread, the number of qubits generally scales with the dataset size (though this is not strictly fixed). For example, for image data with millions of real-valued variables, one would need very large circuits to encode all the information.

The second is amplitude encoding. This technique maps classical data into the amplitudes of a quantum state, offering an exponential reduction in memory, since $M$  variables could be encoded using only $\log_2(M)$ qubits. However, preparing a quantum state that precisely represents the data generally requires an exponential number of operations. This is the same fundamental data-loading problem discussed earlier, for example, in the context of Grover’s algorithm.

This challenge also broadly impacts all quantum linear algebra-based QML methods (not discussed here).\cite{biamonte2017quantum} In these approaches, the apparent exponential speed-up is inherited from the quantum speed-up of algorithms such as the HHL\cite{harrow2009quantum}. But the efficiency of these methods hinges critically on the ability to load classical data into quantum states efficiently, i.e., avoiding the exponential overhead that can negate the advantage.\cite{tang2021quantum,tang2022dequantizing}

\section{Conclusion}
\label{s:conclu}

We discussed several possible applications of quantum computing in the field of genomics, with a particular focus on the roadblocks and challenges that were typically neglected in previous literature.
Note that we focus exclusively on the high-level theoretical challenges associated with the ideal execution of the proposed quantum algorithms in the context of genomics. Additional low-level issues arising from hardware noise or problem-specific implementations are beyond the scope of this discussion.

We began by discussing the conceptual problems related to the use of Grover's algorithm for database search, a task directly connected to alignment.
We showed that the hoped-for, and usually claimed, quadratic advantage vanishes under the realistic assumption of the lack of a QRAM. We also discussed, for the first time in this context, some approximate workarounds, but without any guarantee of accuracy for this important task.

While these concepts are becoming generally known within the quantum information community, we believe it is important to make them known and explain them also to the computational biology community interested in quantum algorithms.
According to current consensus, informed by the foreseeable and realistic characteristics of a future fault-tolerant quantum computer,\cite{babbush:2021} quantum algorithms will likely be hardly useful for so-called big data problems.\cite{hoefler2023disentangling}

Hopes for quantum advantage are connected with the solution of problems that are harder than linear in the dataset size. For this reason, we turned our attention to genomics problems that contain combinatorial optimization tasks.
These are problems that require exponential time, with respect to the input size, to be solved exactly.
However, not all combinatorial optimization problems that are hard in theory are equally difficult in practice.\cite{abbas2024challenges}
We discuss how standard complexity theory may be misleading when guiding us in real-world cases, as sometimes polynomial-time approximation methods are effective in practice even for NP-hard problems.

Genomics problems are inherently affected by errors, which motivates the use of heuristic methods.\cite{kirkpatrick1983optimization,albash2018adiabatic}
However, even in this case, they should not be applied blindly.  
To fit the binary `language' of quantum algorithms, the native optimization problems need to be recast in a mathematical format called QUBO. However, this typically introduces polynomial overheads, which are further amplified if the QUBO problems are embedded into hardware with limited qubit connectivity.\cite{konz2021embedding} In practice, this means that the translated optimization problem to be solved by the quantum computer is significantly harder than the native one.
Given that exponential advantages in runtime are not expected, this introduces a challenge that has never been thoroughly discussed in this application space.

We suggest that problems amenable to quantum speedup need to be carefully identified a priori. In particular, they should correspond mathematically to models that are considered good benchmarks for quantum optimization, because (i) they become hard at much smaller sizes, and (ii) conventional approximate methods perform comparatively worse.\cite{abbas2024challenges,koch2025quantum}
Additionally (iii), they should be mapped into a QUBO problem with minimal overhead.

Among these, we identify the following optimization problems: Max-Cut, Maximum Independent Set, Knapsack, and Quadratic Assignment Problems to be relevant for (or possibly connected to) genomics. Given their mathematical formulations, they could be relevant for problems such as GWAS and gene prioritization. Further, the quantum-enhanced Markov chain algorithm,\cite{layden2023quantum,mazzola2024quantum}, a new quantum algorithm devised to speed up Monte Carlo simulations, can be beneficial for accelerating inference tasks.
It would also be interesting to discover if some genomics tasks admit a formulation as Low Autocorrelation Binary Sequence problems, given that there is some experimental evidence of a scaling advantage using QAOA\cite{shaydulin2024evidence}.  Examples of application of the LABS problem in genomics may include gene selection for cancer subtype classification where binary feature selection can benefit from low autocorrelation to reduce redundancy, and guide RNA (gRNA) design where binary encoding of candidate sequences could be optimized to minimize off-target effects.

Possible quantum advantage will need to be assessed empirically and on a case-by-case basis.
In the absence of sufficiently large hardware capable of directly solving problems of interest, evidence of advantage can be only inferred from runtime scaling analysis. Namely, by estimating the asymptotic quantum runtime by repeating the process at several problem sizes. In this case, we stress the importance of conducting such experiments in an accurate and accepted way, so as not to create artifacts leading to false claims of advantage.\cite{ronnow2014defining,albash2018demonstration}

Finally, we consider quantum machine learning.
In this case, we distinguish between QML algorithms that inherit their theoretical speedup from enhanced quantum linear algebra core functions\cite{biamonte2017quantum} and heuristic algorithms that feature parametrized quantum circuits.
The former approach suffers from the same limitations we discussed regarding the data loading problem.\cite{tang2019quantum,tang2021quantum,tang2022dequantizing} The latter are a suite of heuristic methods that have yet to be proven effective on real-world datasets. Moreover, these methods suffer from critical trainability isses\cite{schuld2022quantum,cerezo2022challenges}
All in all, QML could be a possible path toward quantum advantage if the community understands which kinds of datasets (i.e., with given symmetry, topology, or structure) are amenable to quantum speedup, and, if architectures that are not affected by barren plateaus, and beyond classical simulations, exist.
So far, there is no evidence of quantum advantage in ML beyond artificially constructed datasets (cfn. the recent and very instructive Ref.~\cite{bowles2024better}), although
some works show that graph kernels can be successfully used in quantum support vector machines.\cite{bai2015quantum,henry2021quantum,kishi2022graph}

In conclusion, the goal of this manuscript is to clearly discuss the theoretical challenges of commonly proposed quantum computing solutions for genomics.
Our viewpoint is grounded in the latest advancements within the field.\cite{hoefler2023disentangling,abbas2024challenges,koch2025quantum,layden2023quantum}
With this, we are neither overly pessimistic nor optimistic about the synergies between quantum information and genomics. However, it is time for this field to adopt a mature stance and embrace a balanced perspective on this potentially transformative computational platform. We hope that our work can stimulate new directions and be helpful to researchers in carefully planning their numerical experiments.\\

\textit{Acknowledgments.} 
G.M. acknowledges discussions with  Francesco Tacchino, Zoe Holmes, Massimiliano Incudini.
We acknowledge useful feedback on an earlier version of the manuscript from Cristian Micheletti and Frederik Fl\"other.
G.M. acknowledges financial support from the Swiss National Science Foundation (grant PCEFP2\_203455).

\appendix

\section{Myths on quantum  computing}
\label{app:myths}
In this section, we aim to dispel some frequently made claims regarding the power of quantum computing. Some of these points may be obvious, while others may be less so. We believe this discussion may be particularly useful for non-expert readers.

\textbf{Claim: Quantum gates are faster than classical gates. }Quantum digital gates cannot operate faster than the classical electronics that controls them\cite{krantz2019quantum}. Moreover, in the fault-tolerant quantum era, each logical qubit will be encoded in several physical qubits, and each logical gate will require multiple elementary operations on the individual physical qubits. As a result, the effective quantum clock frequency will be orders of magnitude lower than that of conventional classical processors\cite{fowler2012surface,gidney2019efficient}.\\

\textbf{Claim: A single qubit can store an arbitrarily large amount of information.}
This claim stems from the fact that a qubit state is defined by its coefficients, 
$\alpha$ and $\beta$, which are complex numbers. In principle, the binary expansion of either coefficient could contain an infinite number of digits, suggesting infinite information content.

However, the same argument applies to any classical analog method of storing information. For instance, one could imagine encoding a telephone number by cutting a string to a precise length such that its measurement in millimeters reproduces the exact decimal expansion of the number. Clearly, this is impractical, as such an encoding would be highly susceptible to uncontrollable errors in both preparation and readout. This is precisely why digital computing replaced analog machines.

In the case of qubits, the only possible measurement outcomes are binary, either `0' or `1'. Extracting information about the amplitudes 
$\alpha$ and $\beta$ with arbitrary precision requires performing many measurements in different bases. Therefore, it is impossible to retrieve the exact values of the coefficients in a single shot.\\ 

\textbf{Claim: Quantum algorithms solve a problem by trying all possible inputs in superposition.}
This claim, once again, neglects the fundamental issue of how to extract information from a quantum state through measurements.

It is easy to prepare a quantum state defined on $n$ qubits (all initialized to zero) that contains an equal superposition of all $2^n$ basis states using just one layer of Hadamard gates:
\begin{equation}
\frac{1}{\sqrt{2^n}} \sum_{x=0}^{2^n-1} |x\rangle
\end{equation}

Suppose that all these binary strings are valid inputs to our problem. The task is to find the unique bitstring that satisfies a given set of logical propositions. In simpler terms, we assume that one string is the solution (which we need to find) to an optimization problem.

We imagine having access to a quantum function (oracle) that requires an ancillary qubit. The ancillary qubit, initially prepared in the state $|0\rangle$, flips to $|1\rangle$ only if the input $x$ is the solution to our problem. The process reads as:
\begin{equation}
 \frac{1}{\sqrt{2^n}} \sum_{x=0}^{2^n-1} |x\rangle |0\rangle \rightarrow  \frac{1}{\sqrt{2^n}} \sum_{x=0}^{2^n-1} |x\rangle |f(x)\rangle
\end{equation}
where $f(x) = 1$ if $x$ is the solution $x_\textrm{sol}$, and $f(x) = 0$ otherwise.

We would be tempted to claim victory by asking to find the $n$-qubit register associated with the value of 1 in the ancilla qubit, $|x_\textrm{sol}\rangle | 1 \rangle$ However, this is not possible, and at this stage, the only way to extract the information is to measure the register. Given that all components have the same amplitude, we would obtain a random bit string readout, and the ``wanted'' readout would appear only with a vanishing probability, $1/\sqrt{2^n}$.

At this point, we would have the same efficiency as sampling random input states and checking whether they are solutions to our problem.
The key step to make this rough idea work is to find a way to amplify the wanted component in such a way as to enhance its probability of being read out.
Therefore, the claim overly trivializes the behavior of quantum algorithms (see main text).\\ 

\textbf{Claim: Quantum advantage is directly related to entanglement. }
One may be tempted to claim that any operation producing a sufficiently large entangled state results in a quantum state that is not amenable to classical simulation, and thus satisfies a necessary condition for quantum advantage.
However, if a circuit consists only of quantum gates from the Clifford set, it can be efficiently simulated classically in polynomial time.
Here, by simulation, we mean that it is possible to sample from the distribution of measurement outcomes.\cite{gottesman1998heisenberg}
A current area of research is the theoretical investigation of how many non-Clifford gates—sometimes referred to as magic—are necessary to produce states that are strictly non-classically simulatable.\cite{PhysRevA.71.022316,PhysRevA.86.052329,Haug_2023}
Given that Clifford operations can produce entangled states (for instance, a GHZ state can be prepared using only Hadamard and \textsf{CNOT} gates starting from the `00..0' state) entanglement alone is not a necessary and sufficient condition for quantum advantage.

\section{Computing in quantum superposition}
\label{app:basics}
\subsection{Qubits}

The elementary logic unit of quantum computing is the qubit, which, unlike a conventional bit, can exist in a superposition of both basis states $|0\rangle$ and $|1\rangle$.\cite{nielsen:2000} Mathematically, these basis states 
form an orthonormal basis for the complex vector space $ \mathbb {C}^2$, such that a general qubit state, or wavefunction, reads $| \psi \rangle = c_0 |0\rangle + c_1 |1\rangle $. The complex coefficients, $c_0$ and $c_1$ associated with these states must satisfy the normalization condition, $|c_0|^2+|c_1|^2 = 1$ which derives from the laws of quantum mechanics. These coefficients, squared, represent the probability of measuring the qubit in either the state `0' or `1', therefore their sum, squared, must equal to one.

Quantum computing with a single qubit involves applying a unitary transformation, $U$ to the initially prepared state to produce a final state, $| \psi \rangle \rightarrow U | \psi \rangle = | \psi ' \rangle = c_0' |0\rangle + c_1' |1\rangle  $.
The computation must conclude with a measurement process that collapses the wavefunction onto one of the basis states. 
For instance, now the wavefunction will collapse on the state `0' with probability  $|c_0'|^2$ and on `1' with probability  $|c_1'|^2$

Single-qubit gates are mathematically defined as unitary transformations in the $ \mathbb {C}^2$
  space, and in practice they are represented by $2 \times 2$ matrices.
   Two-qubits gates are represented  by a $2^2 \times 2^2$
unitary matrices, while three-qubit gates  by  $2^3 \times 2^3$ objects, and so on. Each hardware implementation has different capabilities regarding specific gate sets, but canonical gate sets are commonly adopted in quantum information textbooks.\cite{nielsen:2000} 
  In this Section we will use for illustrative purposes just a few gates: the NOT gate, $\textsf{X}$, which acts on the basis states as $\textsf{X}|0\rangle = |1\rangle $ and $\textsf{X}|1\rangle = |0\rangle $,
  the gate, $\textsf{Z}$, which acts on the basis states as $\textsf{Z}|0\rangle = |0\rangle $ and $\textsf{Z}|1\rangle = -|1\rangle $, and their controlled versions,
  the \textsf{CNOT} (controlled-\textsf{X}) gate and the \textsf{C-Z} (controlled-\textsf{Z}) gate.
   See the Appendix\ref{app:gates} for a formal definition of all the gates used in this section, and their matrix representation. Due to the linearity of quantum mechanics, it is sufficient to define the action of the gates on the basis vectors, their action on the general state reads, for instance, $\textsf{Z}| \psi \rangle = c_0 |0\rangle  - c_1 |1\rangle $.
  For universal computation, the unitary transformations in the gate set must approximate any other unitary transformation arbitrarily well. If these conditions are met, quantum circuits can be constructed, with the assurance that the unitary operations can be executed with a controllable error. 
  However, creating an arbitrary multi-qubits unitary operation is not guaranteed to be efficient, meaning it may scale exponentially with the number of qubits, $n$. We will return to this important point later.

  \subsection{Quantum circuits and superposition}

To achieve practical computation, one must work with more than one qubit. In this case, the total computational space increases exponentially with the number of qubits, leading to the popular statement that a quantum computer can achieve exponential memory compression. For example, the basis states of a $n=3$ qubit computational space are: $|000\rangle$, $|001\rangle$, $|010\rangle$, $\cdots$, $|111\rangle$, such that a general three qubit state is defined via$2^3 = 8$ complex coefficients, and reads
$| \psi \rangle = c_{000}|000\rangle+ c_{001} |001\rangle +c_{010} |010\rangle + \cdots + c_{111}|111\rangle $, where the sum of the coefficients squared, $|c_j|^2$, is one.

In reality, the power of quantum computation lies not just in memory compression but in the ability to apply a function to each component of the wavefunction simultaneously, due to superposition. Classically, to determine the action of a function 
$U$ on each possible bitstring, one would need to apply that $2^n$
  times, i.e., once for each bitstring.

\begin{figure}[t]
    \centering
    \includegraphics[width=0.9\columnwidth]{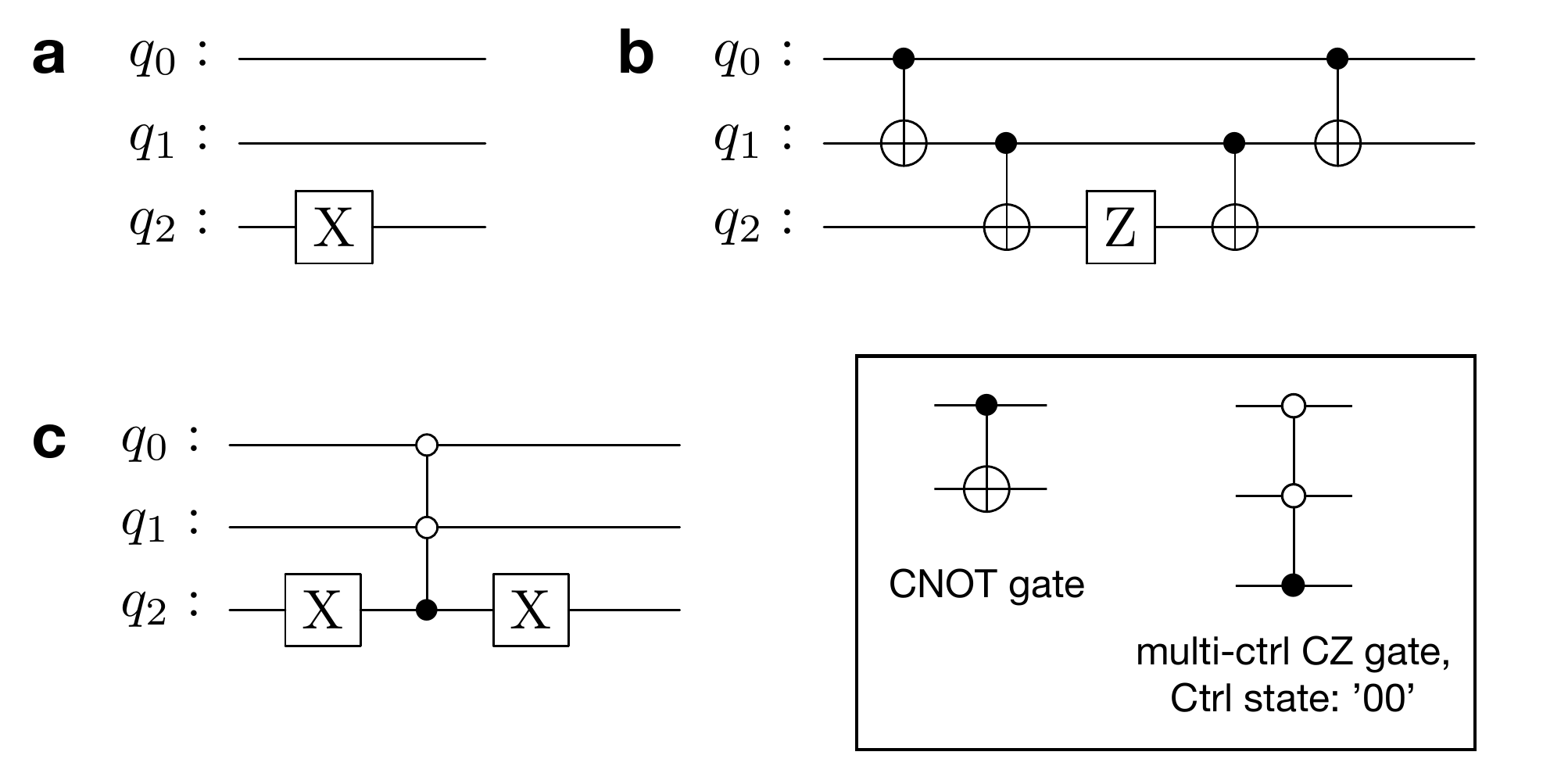}
    \caption{Example of circuits representing the unitary functions described in the main text. (a) A circuit that flips the value of the leftmost bit; (b) a circuit that changes the sign of the amplitudes of the components with an odd number of '1'; (c) a circuit that changes the sign of the amplitude of the component corresponding to the '000' state. The multi-qubit gates used are defined in Appendix.
    }
    \label{fig:examples_circuits}
\end{figure}

In the quantum setup, the function $U$, which operationally is represented by a particular gate or sequence of gates, is applied to all $2^n$
  components of the wavefunction at once, thanks to the linearity of quantum mechanics.
For instance, let's consider the simplest function: flipping the leftmost bit in our previous 3-qubit example. To accomplish this, one would apply the $\textsf{X}$ gate to the first qubit while leaving the other qubits unchanged. Mathematically, since the full unitary operation is defined over the three-qubit space, 
$U$ would be the tensor product of the  $\textsf{X}$ gate acting on the first qubit and identity,  $\textsf{I}$, operations on the second and third qubits, namely  $U =  \textsf{I}\otimes  \textsf{I} \otimes\textsf{X}  $.
Assuming, little-endian convention about bit-ordering, $U | 000 \rangle = | 100 \rangle  $ and so on. The action of $U$ on the state defined above produces the new state
$| \psi' \rangle = c_{000}|100\rangle+ c_{001} |101\rangle +c_{010} |110\rangle + \cdots + c_{111}|011\rangle $. For instance, the new coefficient multypling the state $|000\rangle$ is the one previously `attached' to $|100\rangle$, so $c_{000}' = c_{100}$. The corresponding, simple, circuit is shown in Fig.\ref{fig:examples_circuits}.a.

As a second example, we consider the following ``function'': detecting whether a bitstring has an even or odd number of `1's, and flipping the sign of the corresponding coefficients in the latter case. A quantum circuit that corresponds to this \textit{parity check} unitary is depicted in Fig.\ref{fig:examples_circuits}.b. This unitary acts on the superposition of bitstrings, simultaneously flipping the signs of all the relevant coefficients.
This unitary yields the output state 
$ U|\psi \rangle = | \psi' \rangle = c_{000}|000\rangle - c_{001} |001\rangle - c_{010} |010\rangle + c_{011} |011\rangle + \cdots - c_{111}|111\rangle $
A final example is a unitary that selectively flips the sign of the coefficient for only a specific target bitstring, such as the $|000\rangle$ state. The corresponding circuit is shown in Fig.\ref{fig:examples_circuits}.c, and the resulting state would be $ U|\psi \rangle = | \psi' \rangle = -c_{000}|000\rangle + c_{001} |001\rangle + c_{010} |010\rangle +  \cdots + c_{111}|111\rangle $.

At this point, some readers may raise the concern that these examples seem not useful at all, while others might already glimpse a connection with the task of database search. Both perspectives are valid. Indeed, some of these functions are useless in isolation, but they serve as building blocks for more complex and useful algorithms.
For instance, the third example can be understood as an oracle that \emph{marks} the 
`000' string. Oracles are essential components of many algorithms, such as Grover's algorithm.\cite{nielsen:2000,grover1996fast,boyer1998tight} They act as black boxes, performing specific operations.

In Grover's algorithm, each oracle application can be viewed as a database query. 
The algorithm's asymptotic speed-up is measured by the number of times the database must be consulted to find the desired entry. Therefore, the quadratic speed-up of Grover's algorithm, often cited as a quantum advantage for database search, pertains to the number of oracle queries rather than the total number of elementary operations.
This concept is crucial and will be explored in greater detail later.

Returning to our last two examples ( Fig.\ref{fig:examples_circuits}.b and Fig.\ref{fig:examples_circuits}.c), marking a state by changing the sign of its amplitudes would not result in an effective search algorithm, as it does not alter the measurement outcome in the computational basis. Recall that the amplitude squared represents the probability of measuring a specific bit string.

The common theme of successful quantum algorithms is to perform operations in such a way that the target bit string—whether it represents the data to be found or the solution to an optimization problem—is measured with high probability.
In Appendix~\ref{app:groverexample} we see that the third function defined above is actually Grover's oracle in the particular case of `marking' a database entry encoded with the '00' bitstring, but an additional unitary is required to amplify its amplitude.

\section{Quantum gates}
\label{app:gates}

This Appendix provides the definitions of the quantum gates used in this work. The matrix representations are given in the computational basis $\{|0\rangle, |1\rangle\}$ for single-qubit gates and their tensor product extensions for multi-qubit gates.

The Pauli-X gate is a single-qubit gate represented by the following $2 \times 2$ matrix:
$$
\textsf{X} = \begin{pmatrix}
0 & 1 \\
1 & 0
\end{pmatrix}
$$
Its action on the computational basis states is:
$$
\textsf{X}|0\rangle = |1\rangle , \quad
\textsf{X}|1\rangle = |0\rangle
$$

The Pauli-Z gate is a single-qubit gate represented by the following $2 \times 2$ matrix:
$$
\textsf{Z} = \begin{pmatrix}
1 & 0 \\
0 & -1
\end{pmatrix}
$$
Its action on the computational basis states is:
$$
\textsf{Z}|0\rangle = |0\rangle, \quad
\textsf{Z}|1\rangle = -|1\rangle
$$

The Hadamard gate is a single-qubit gate represented by the following $2 \times 2$ matrix:
$$
\textsf{H} = \frac{1}{\sqrt{2}} \begin{pmatrix}
1 & 1 \\
1 & -1
\end{pmatrix}
$$
Its action on the computational basis states is:
\begin{eqnarray}
\nonumber
\textsf{H}|0\rangle = \frac{1}{\sqrt{2}}(|0\rangle + |1\rangle) \\
\nonumber
\textsf{H}|1\rangle = \frac{1}{\sqrt{2}}(|0\rangle - |1\rangle)
\end{eqnarray}

The Controlled-NOT (\textsf{CNOT}) gate is a two-qubit gate. The first qubit is the control qubit, and the second qubit is the target qubit. The \textsf{CNOT} gate flips the state of the target qubit if and only if the control qubit is in the state $|1\rangle$. Its $4 \times 4$ matrix representation in the basis $\{|00\rangle, |01\rangle, |10\rangle, |11\rangle\}$ is:
$$
\textsf{CNOT} = \begin{pmatrix}
1 & 0 & 0 & 0 \\
0 & 1 & 0 & 0 \\
0 & 0 & 0 & 1 \\
0 & 0 & 1 & 0
\end{pmatrix}
$$
Its action on the computational basis states is:
\begin{eqnarray}
\nonumber
\textsf{CNOT}|00\rangle = |00\rangle \\ \nonumber
\textsf{CNOT}|01\rangle = |01\rangle \\ \nonumber
\textsf{CNOT}|10\rangle = |11\rangle \\ \nonumber
\textsf{CNOT}|11\rangle = |10\rangle \nonumber
\end{eqnarray}

The multicontrolled-CZ gate with 3 qubits, where the first two qubits are the control qubits and the third is the target qubit. In this case we define the control string to be `00'. This applies a Pauli-Z gate to the target qubit only when the first two control qubits are in the state $|00\rangle$. The $8 \times 8$ matrix representation in the basis $\{|000\rangle, |001\rangle, |010\rangle, |011\rangle, |100\rangle, |101\rangle, |110\rangle, |111\rangle\}$ is:
$$
\textsf{CZ}_{00}^{(1,2|3)} = \begin{pmatrix}
-1 & 0 & 0 & 0 & 0 & 0 & 0 & 0 \\
0 &- 1 & 0 & 0 & 0 & 0 & 0 & 0 \\
0 & 0 & 1 & 0 & 0 & 0 & 0 & 0 \\
0 & 0 & 0 & 1 & 0 & 0 & 0 & 0 \\
0 & 0 & 0 & 0 & 1 & 0 & 0 & 0 \\
0 & 0 & 0 & 0 & 0 & 1 & 0 & 0 \\
0 & 0 & 0 & 0 & 0 & 0 & 1 & 0 \\
0 & 0 & 0 & 0 & 0 & 0 & 0 & 1
\end{pmatrix}
$$
Its action on the computational basis states is:
\begin{eqnarray}
\nonumber
&\textsf{CZ}_{00}^{(1,2|3)}|000\rangle = -|000\rangle \\ \nonumber
&\textsf{CZ}_{00}^{(1,2|3)}|001\rangle = -|001\rangle \\ \nonumber
&\textsf{CZ}_{00}^{(1,2|3)}|010\rangle = |010\rangle\\  \nonumber
&\textsf{CZ}_{00}^{(1,2|3)}|011\rangle = |011\rangle \\ \nonumber
&\textsf{CZ}_{00}^{(1,2|3)}|100\rangle = |100\rangle \\ \nonumber
&\textsf{CZ}_{00}^{(1,2|3)}|101\rangle = |101\rangle \\ \nonumber
&\textsf{CZ}_{00}^{(1,2|3)}|110\rangle = |110\rangle \\ \nonumber
&\textsf{CZ}_{00}^{(1,2|3)}|111\rangle = |111\rangle
\end{eqnarray}
In general, for a multicontrolled gate with control qubits $c_1, c_2, \dots, c_n$ and a target qubit $t$, conditioned on a control string $s = s_1s_2\dots s_n$, the target qubit is acted upon by a single-qubit gate $U$ if and only if the control qubits are in the state $|s_1\rangle|s_2\rangle\dots|s_n\rangle$. For the controlled-CZ gate with control string '00', the single-qubit gate $U$ is the Z gate.

\section{Grover algorithm for a toy database search without QRAM}
\label{app:groverexample}

\begin{figure*}[ht]
    \centering
    \includegraphics[width=0.9\textwidth]{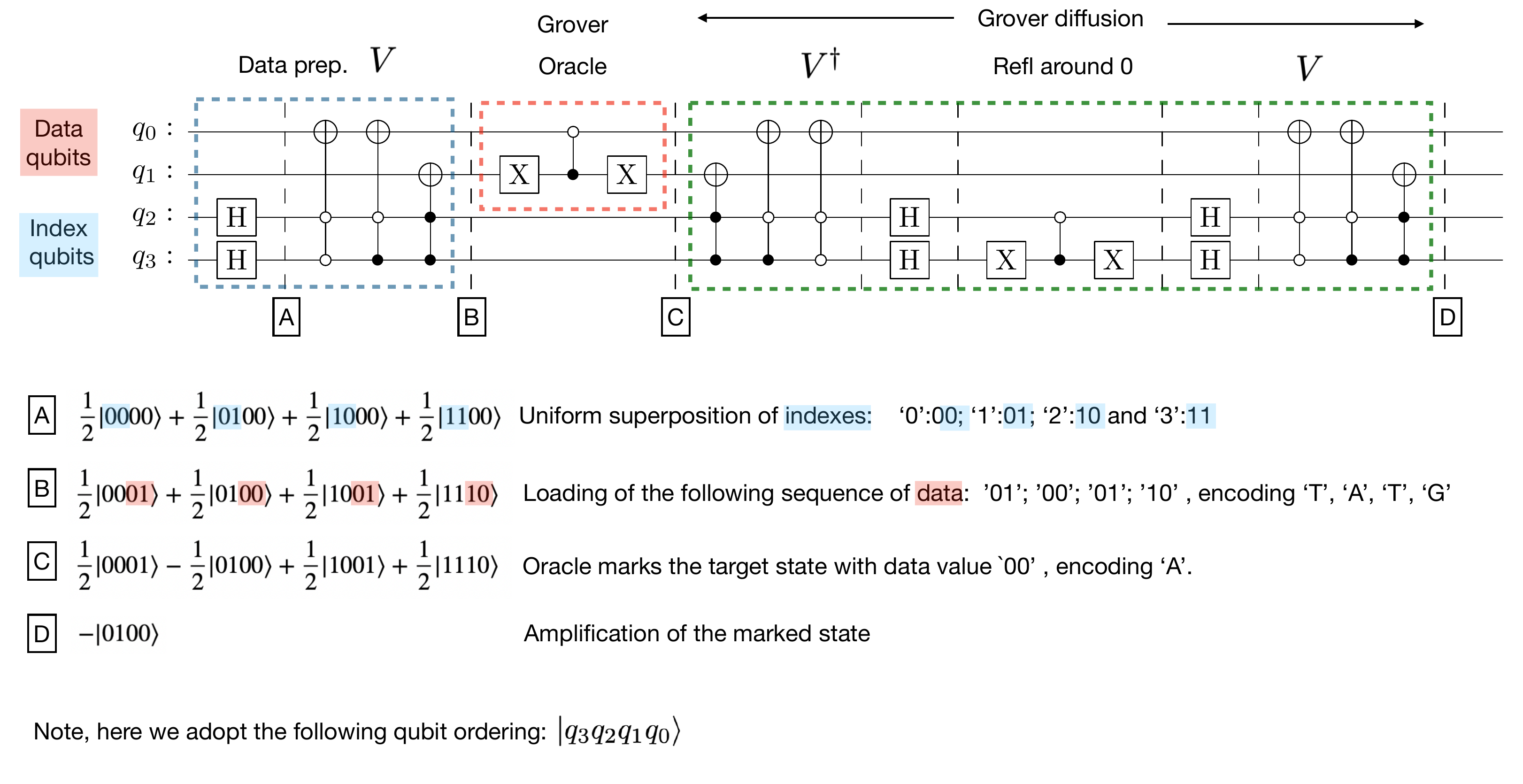}
    \caption{Grover's algorithm with index and data encoding.
    }
    \label{fig:examples_grover}
\end{figure*}

Suppose we have the following list of elements, $\{T, A, T, G\}$, and we want to find the position of the element A using Grover's algorithm. To encode the data, we need two qubits (see main text), but we also require an additional qubit register to store the indices  $\{0, 1, 2, 3\}$  of the elements. Therefore, we use four qubits: two for encoding the indices and two for encoding the data.
A generic quantum state is the tensor product of the two registers $|q_3 q_2 \rangle \otimes |q_1 q_0 \rangle =  |q_3 q_2 q_1 q_0 \rangle$

Most textbook examples of Grover's algorithm initialize the list using a uniform superposition of strings. However, in this case, the process is more nuanced because we need to encode both the indices and the data together, meaning the data must be tied to the corresponding indices.
Our goal is to initialize the quantum state in such a way as to encode the full input
$\{(0,T), (1,A), (2,T), (3,G)\}$.

The first step is to create a uniform superposition of indices in the bottom register (see Fig.~\ref{fig:examples_grover})
$1/2( |00\rangle + |01\rangle + |10\rangle + |11\rangle) \otimes |00\rangle $.
Then we suitably entangle each data entry with its corresponding index:
$1/2( |00\rangle \otimes |'T'\rangle + |01\rangle \otimes |'A'\rangle + |10\rangle \otimes |'T'\rangle + |11\rangle \otimes |'G'\rangle) $, where we used the data values A,T,G, with a bit of abuse of notation, to better distinguish the two conceptually different bit positions.
This is achieved through the state preparation unitary $V$, which initializes the entire qubit register from the initial 
$|0000\rangle$ state.
Notice that in Fig.~\ref{fig:examples_grover} the bottom register (qubits $q3,q2$) encodes the index, while the top register (qubits $q1,q0$) encodes the data in binary format. We adopt the little-endian convention such that the bottom qubit are the leftmost in the bitstring representation.

In this example, we use a sequence of multicontrolled \textsf{CNOT} gates. For instance, to entangle the `T' entry with the first index position `00', we place a multicontrolled \textsf{CNOT} gate acting on the first data qubit, $q_0$, controlled by the index string `00'. Similarly, to entangle the last entry `G' with the position `11', we use a multicontrolled \textsf{CNOT} gate acting on the second data qubit, controlled by the index string `11'.
To entangle the data entry `A' with the index `01' we simply do nothing because our encoding assigns the `00' value to A, so no bit has to be flipped.
Multi-controlled gates can be compiled using \textsf{CNOT} gates\cite{nielsen:2000}. In Fig.~\ref{fig:examples_grover} we avoid doing for the sake of visualization.
Notice that here we use a purely quantum mechanical feature of controlling the execution of a function, in superposition.

For all practical purposes, this section could conclude here. The data loading unitary 
$V$ necessarily contains order of 
$N$ gates, given that, for each index, we had to design a suitable multicontrolled \textsf{CNOT} gate to entangle the corresponding data.\cite{park2019circuit}

If the data sequence is pseudo-random, lacks any regular periodicity, or cannot be mathematically modeled, each entry must be loaded individually using an order of $N$ operations. This means that the data loading phase already negates the quadratic speed-up.

For the sake of example, we proceed further with the Grover circuit. The Grover operator is divided into two steps that need to be iterated. The crucial point is that the number of iterations required to amplify the amplitude of the desired state $|\textrm{index},\textrm{data} \rangle $ is 
$\sqrt{N}$ rather than $N$

The first operation is usually called `reflection' around the marked state' and is typically performed by an oracle in textbooks. In our specific implementation, since we encode both the index and the data, we can explicitly provide a circuit for this oracle. The marked state is contained in the `data register', and in our encoding, it corresponds to the `00' state. This information is known to us a priori. Therefore, the circuit implements the function discussed in the main text (see Fig.~\ref{fig:examples_circuits}.c), i.e., a unitary that flips the sign of the components with `00' in the data register.

The second operator is called diffusion or reflection around the initial state. If the initial state is simply a uniform superposition of all four qubit states, the circuit is straightforward. However, in this case, it is not, since we need to use the reverse unitary, $V^\dag$, that prepared the entangled state, apply a reflection around `00' in the index register, and then apply $V$ again.

In this simple example, the a priori success probability of finding the correct entry is 25$\%$, given that there are just four elements. This is a limiting case in which the Grover algorithm returns the correct element in just one iteration. The final state produced by the circuit in Fig.~\ref{fig:examples_grover} is indeed 
$|0100\rangle$, that encodes the value $('1'=01,'A'=00)$.

The total runtime of this implementation would scale as $N\sqrt{N}$, because every Grover step also requires a depth $N$ subcircuit, thus realizing a quantum slowing-down compared to the brute force classical approach.

\bibliography{biblio}

\end{document}